\documentclass[pdflatex,sn-mathphys-num]{sn-jnl}

\usepackage{graphicx}%
\usepackage{multirow}%
\usepackage{amsmath,amssymb,amsfonts}%
\usepackage{amsthm}%
\usepackage{mathrsfs}%
\usepackage[title]{appendix}%
\usepackage{xcolor}%
\usepackage{textcomp}%
\usepackage{manyfoot}%
\usepackage{booktabs}%
\usepackage{algorithm}%
\usepackage{algorithmicx}%
\usepackage{algpseudocode}%
\usepackage{tikz}%
\usetikzlibrary{arrows.meta,positioning,calc,bending}%
\graphicspath{{./}}%
\definecolor{rred}{RGB}{200,40,40}%
\definecolor{flowblue}{RGB}{0,102,204}%
\definecolor{transblue}{RGB}{100,150,255}%
\definecolor{myorange}{RGB}{230,126,34}%
\definecolor{netpurple}{RGB}{120,70,160}%

\newcommand{\taux}{\tau_x}
\newcommand{\tauy}{\tau_y}
\newcommand{\Retau}{\mathrm{Re}_{\tau}}

\newcommand{\what}{\widehat{w}}
\newcommand{\tplus}{t^{+}}

\theoremstyle{thmstyleone}

\theoremstyle{thmstyletwo}

\theoremstyle{thmstylethree}

\begin{document}

\title[Offline accuracy is not enough]{Offline accuracy is not enough:
closed-loop instability and stabilisation of a wall-sensor neural estimator in
opposition control}

\author*[1]{\fnm{Giorgio Maria} \sur{Cavallazzi}}\email{giorgio.cavallazzi@city.ac.uk (ORCID: 0000-0001-7529-3256)}
\author[1]{\fnm{Miguel} \sur{P\'erez Cuadrado}}
\author[1]{\fnm{Alfredo} \sur{Pinelli}}

\affil*[1]{\orgdiv{Department of Engineering, School of Science and Technology},
  \orgname{City St George's, University of London},
  \orgaddress{\city{London}, \country{United Kingdom}}}

\abstract{Opposition control reduces skin-friction drag by opposing the wall-normal velocity on a near-wall detection plane, but the detection-plane velocity it requires is not available from wall-mounted sensors. Wall data can reconstruct inner-flow quantities accurately when assessed offline on a fixed flow state, and we ask whether such a reconstructed field can instead serve as a live surrogate sensor inside the feedback loop. We train a recurrent estimator to infer the detection-plane velocity from the two wall-shear-stress components in opposition-controlled turbulence. Offline it performs extremely well, reaching a correlation of 0.99 and near-unity coherence across the energetic scales; yet the same estimator fails in closed loop, decorrelating from the true field within a few viscous time units as the control collapses. The failure is not one of accuracy but of distribution shift induced by the controller itself: small closed-loop errors carry the flow off the attractor represented in the training data, while unresolved high-wavenumber errors enter through the wall boundary condition and return as out-of-distribution inputs. Standard remedies such as low-pass filtering and exponential averaging only delay numerical breakdown while accelerating decorrelation. Stable wall-only control is recovered by imposing spectral consistency on the deployed actuation and retraining the estimator on its own closed-loop data, giving a controller that holds much of the drag reduction of ideal opposition control from wall quantities alone. The obstacle is not whether the near-wall flow can be reconstructed offline, but whether that reconstruction stays dynamically consistent when allowed to modify the flow it senses.}

\maketitle

\section{Introduction}\label{sec:intro}

Skin-friction drag accounts for a large share of the energy spent moving vehicles and fluids through pipes, so that even modest reductions carry a substantial economic and environmental payoff~\cite{GadelHak2000,spalart2011dragreduction}. 
A great deal of effort has consequently gone into manipulating wall-bounded
turbulence to lower the drag it generates, and the active flow control strategies divide broadly into
predetermined forcing, which imposes a fixed actuation independent of the
instantaneous flow (spanwise wall oscillation and streamwise-travelling waves
of spanwise velocity \cite{quadrio2009stw,touber2012spanwise,gatti2016stwhighre},
or uniform blowing and suction
\cite{kametani2011blowingsuction,kametani2015blowingsuctionTBL,hasegawa2011dissimilar}),
and reactive control, which senses the flow and responds to it in closed loop
\cite{brunton2015closedloop,chung2011effectiveness}. The two families trade
robustness for efficiency: predetermined schemes need no sensing but spend power
continuously, whereas reactive schemes can in principle be far more economical
\cite{marusic2021energypath} because they act only where and when the flow
requires it. This efficiency arises  because the skin friction is set almost
entirely by the near-wall Reynolds shear stress \cite{fukagata2002fik}, so a
controller that targets the quasi-streamwise vortices responsible for that stress
can produce a large effect with a small, well-placed actuation.

The canonical reactive law is opposition control \cite{choi1994opposition}: a
blowing-and-suction velocity is imposed on the wall equal and opposite to the
wall-normal velocity measured on a detection plane a short distance above it,
building a virtual wall that screens the surface from the overlying vortices
\cite{hammond1998oppositionmech} and lowering the drag by twenty to
twenty-five percent at low Reynolds number. The mechanism and its limits have been
mapped across viscous scales \cite{chang2002viscous} and between channel and
boundary-layer geometries \cite{stroh2015oppositioncomparison,pamies2007opposition}. The idea also connects 
to earlier, model-free strategy of selectively sucking out bursting events \cite{gadelhak1989selectivesuction}. More elaborate
feedback laws derived from optimal and suboptimal control theory
\cite{lee1998suboptimal,fukagata2004suboptimal} push the achievable drag reduction
towards the bound set by full-state predictive control \cite{bewley2001predictive}
and by net-power arguments \cite{fukagata2009netpower}. 

There has also been some experimental progress towards reactive wall control based on blowing and suction. Early laboratory tests used a wall-normal jet from a piston-type actuator to oppose sweep events, showing that appropriately timed opposition-type actuation can modify the burst signature of a turbulent boundary layer, while also avoiding the need for a fully real-time distributed implementation because of the high-frequency response requirements~\cite{RebbeckChoi2006}. More recently, large-scale reactive opposition-control prototypes have used wall-mounted sensing and localised blowing to attenuate large-scale motions and modify the wall stress in turbulent boundary layers~\cite{dacome2024opposition}. These experiments are essential steps towards practical implementation, but they are not the fully distributed, wall-unit-scale and high-speed form assumed in the canonical numerical formulation of opposition control. In that formulation the controller requires the wall-normal velocity on a detection plane in the buffer layer, or even fuller information about the velocity field away from the wall. This remains the central sensing obstacle: any deployable sensor lives at the wall, whereas the quantity required by the controller does not. 

Even the wall measurements that would have to substitute for that off-wall signal are themselves challenging at the required fidelity.
Techniques based on microelectromechanical sensors, oil-film methods and thermal probes have matured substantially in laboratory settings ~\cite{naughton2002shearstress}, but none yet delivers the spatial resolution and bandwidth that distributed opposition control demands at realistic Reynolds numbers.

Having excluded direct inner-flow measurements, the question is how much of the off-wall flow is encoded in the wall footprint. Partial answers exist, but mainly in an offline reconstruction setting. 
Linear stochastic estimation from the two wall-shear components and the wall pressure is able to reconstruct the
near-wall field, with the fidelity set by the instantaneous wall signal up to $y^{+}\!\approx\!20$ \cite{suzuki2017estimation}. 
The large wall-attached eddies that hold much of the energy and Reynolds stress remain observable from the wall into the logarithmic layer, a result that has been used to argue for wall-based detection and control of those structures \cite{encinar2019wallview}. 
Deep learning has made it possible to sharpen the mapping by reconstructing velocity fields from coarse wall measurements with convolutional networks
\cite{guemes2021wall,guastoni2025fcn}. 
These results prove that wall data carry substantial information about the flow above, but only in the offline sense: the estimator is evaluated on data drawn from the same flow state on which the map was trained. Whether that map survives deployment inside a feedback loop is the question this paper addresses.

Taking wall-based sensing as the baseline constraint,  we learn the map from the wall-shear stresses $(\taux,\tauy)$ to the
detection-plane wall-normal velocity, feeding the estimate to the control law in
place of the true signal. A single wall snapshot under-determines the off-wall field, so the estimator carries temporal context across control cycles through a gated recurrent unit~\cite{cho2014gru}; the same combination of spatial and temporal context proved necessary for learned convection control~\cite{cavallazzi2026convection}.

An alternative paradigm avoids this inference problem entirely. Deep reinforcement learning produces wall-actuation policies that match or exceed opposition  \cite{guastoni2023drl,sonoda2023rl,han2020cnnchannel,lee2023drlchannel}, adapt across Reynolds numbers \cite{varela2022actuators,zhou2025drlhighre}, and target the regeneration cycle rather than the shear stress alone \cite{cavallazzi2024wallcycle}; the broader use of machine learning for flow control is reviewed in \cite{brunton2020mlreview,garnier2021reviewdrl,vignon2023review}. A reinforcement-learning agent sidesteps the sensing problem by construction: it learns on the closed-loop flow it creates, so its training distribution is the controlled attractor and co-evolves with the policy as training proceeds.

The present paper takes a different route, one that keeps the original opposition-control law and asks only whether a supervised estimator, trained offline, can supply the detection-plane velocity it requires. The appeal is interpretability and economy: no closed-loop training, no reward shaping, a fixed and transparent control law. The difficulty is that offline accuracy gives no guarantee of closed-loop stability. This is the familiar failure of a policy trained on one distribution and deployed on another~\cite{ross2011dagger}, and appears equally in learned solvers deployed beyond their training data~\cite{brandstetter2022lpsda,um2020solverloop}.
Here the instability is concentrated in the high-wavenumber band where the wall carries little usable information: once imposed as a boundary condition, small-scale errors feed back into the wall-shear signal and rapidly drive the estimator off its training distribution. Common stabilising devices, such as low-pass filtering and exponential smoothing, delay numerical collapse but accelerate the loss of correlation. We treat the failure as diagnostic rather than as something to be circumvented.

It is found that two ingredients stabilise the loop: a spectral truncation that removes the unstable modes from the deployed field, and a short retraining phase on data collected from the controlled flow itself. No differentiable solver is required: at each cycle, the direct numerical simulation supplies the detection-plane velocity as a label. The spectral constraint keeps the loop stable long enough to collect these data, while retraining moves the estimator onto the distribution it will encounter in deployment. The result is a stable wall-only controller that recovers near-opposition-control drag reduction and settles onto a controlled state very close to that of classical opposition control.
\section{Methods}\label{sec:methods}

\subsection{Flow configuration and opposition control}\label{sec:setup}

The flow obeys the incompressible Navier--Stokes equations,
\begin{equation}
\frac{\partial \mathbf{u}}{\partial t} + (\mathbf{u}\cdot\nabla)\mathbf{u}
 = -\nabla p + \nu\,\nabla^{2}\mathbf{u} + \Pi(t)\,\mathbf{e}_{x},
\qquad \nabla\cdot\mathbf{u}=0,
\label{eq:ns}
\end{equation}
with velocity $\mathbf{u}=(u,v,w)$ along the streamwise, spanwise and
wall-normal directions $(x,y,z)$, kinematic pressure $p$, and a spatially
uniform body force $\Pi(t)$ in the streamwise direction adjusted at every step
to hold the bulk velocity fixed. The domain is periodic in $x$ and $y$; the
bottom is a no-slip wall and the top a free-slip boundary, so that in the
uncontrolled case $\mathbf{u}=\mathbf{0}$ at $z=0$. 

The sensor signals consist of the two wall-shear-stress components,
\begin{equation}
\taux = \nu\,\frac{\partial u}{\partial z}\bigg|_{z=0},
\qquad
\tauy = \nu\,\frac{\partial v}{\partial z}\bigg|_{z=0},
\label{eq:tau}
\end{equation}
from which the friction velocity $u_\tau=\sqrt{\nu\,\partial_z\langle u\rangle|_{0}}$
and the friction Reynolds number $\Retau=u_\tau\delta/\nu$ follow, with $\delta$
the channel half-height and $\langle\cdot\rangle$ a wall-parallel average;
quantities in viscous units carry a $+$ superscript and use $u_\tau$ and $\nu$.
Control enters the problem only through the wall-normal velocity imposed at
$z=0$, a zero-net-mass blowing-and-suction field set by the opposition law.

We consider an open turbulent channel at friction Reynolds number
$\Retau=180$, integrated by direct numerical simulation (DNS) over a domain of
size $L_x\times L_y\times L_z = 10.68\,\delta\times3.2\,\delta\times\delta$,
which in viscous units spans $1922\times576\times180$ wall units. The domain is
discretised on a $N_x\times N_y\times N_z = 256\times256\times100$ grid, uniform
in the two wall-parallel directions with spacings $\Delta x^{+}\approx7.5$ and
$\Delta y^{+}\approx2.3$ and stretched in the wall-normal direction from
$\Delta z^{+}\approx0.4$ at the wall to $\approx3.4$ at the channel centre, a
standard DNS resolution. Time advancement uses a
pressure-correction scheme, implemented in the GPU-accelerated finite-difference
solver CaNS~\cite{costa2018cans,costa2021gpucans} with the cuDecomp adaptive
pencil-decomposition library~\cite{romero2022cudecomp}.

The opposition-control law of Choi, Moin and Kim~\cite{choi1994opposition} sets
the wall blowing-and-suction velocity equal and opposite to the wall-normal
velocity on a detection plane at $z_d^{+}\!\approx\!14$,
\begin{equation}
w(x,y,\,z{=}0) = -\,\alpha\,w(x,y,\,z_d),
\label{eq:oc}
\end{equation}
with the gain $\alpha$ set to unity unless stated otherwise. This is our
reference controller, hereafter \emph{OC}; at the present Reynolds number it
yields a drag reduction of approximately $22.7\%$ relative to the uncontrolled
channel driven at the same flow rate. 

We replace the true detection-plane velocity with the wall-only estimate $\what$, 
\begin{equation}
w(x,y,\,z{=}0) = -\,\alpha\,\what(x,y),
\label{eq:ocbc}
\end{equation}
where $\what$ is the network's estimate of $w(z_d)$ from $(\taux,\tauy)$. The
exact law~\eqref{eq:oc} is recovered when $\what$ equals the true
detection-plane velocity, and the question is how close a wall-only estimator
can come to it.

The numerical implementation closes the control loop as follows. 
The solver is extended with a coupling layer that, once per control cycle,
exports the wall-shear fields to the neural estimator and imposes the resulting
blowing-and-suction velocity as the wall boundary condition, closing the loop
synchronously with the simulation. Sensing is restricted to the bottom wall,
where both shear components are available. The time step is
$\mathrm{d}t=0.0085\,\delta/U_b$, i.e.\ just under $0.1\,t^+$ ($t^+=\nu/u_\tau^2$ is the viscous time unit). The control
is held fixed for $n_{\mathrm{act}}=10$ time steps before being updated, so
that one control cycle roughly equals one viscous time unit.

\subsection{Wall-only estimator and control loop}\label{sec:estimator}

\begin{figure}[tbp]
\centering
\resizebox{0.98\linewidth}{!}{%
\begin{tikzpicture}[every node/.style={font=\footnotesize}, >=Stealth,
    axis/.style={thick,->}]
  \begin{scope}[xshift=-5.6cm, x={(1,-.25,-.25)}, y={(0,1,0)},
                z={(.25,0,-1.25)}, scale=0.42]
    \input{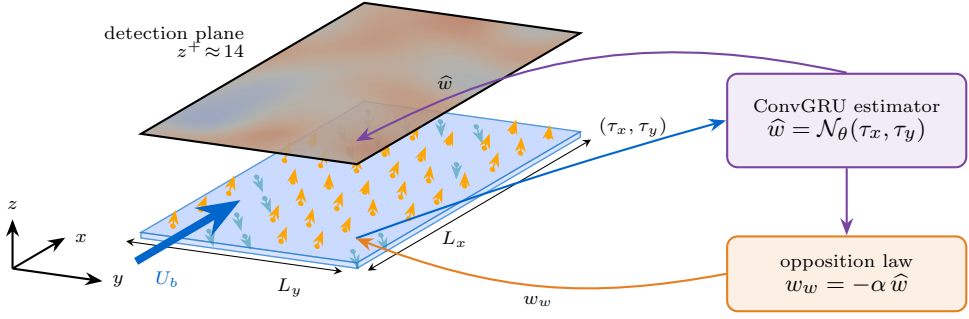}
  \end{scope}
  \node[align=right] at (-6.6,2.75)
        {detection plane\\[-1pt]$z^+\!\approx\!14$};
  \node[draw=netpurple, fill=netpurple!10, thick, rounded corners,
        minimum width=3.2cm, minimum height=1.25cm, align=center]
        (net) at (2.4,1.7)
        {ConvGRU estimator\\[1pt]\scriptsize $\widehat{w}=\mathcal{N}_\theta(\tau_x,\tau_y)$};
  \node[draw=myorange, fill=myorange!12, thick, rounded corners,
        minimum width=3.2cm, minimum height=1.0cm, align=center]
        (law) at (2.4,-0.35)
        {opposition law\\[1pt]\scriptsize $w_w=-\alpha\,\widehat{w}$};
  \draw[->, thick, flowblue]
        (WALL_P) to[bend left=2]
        node[pos=0.68,above,xshift=12pt,yshift=4pt,black]{$(\tau_x,\tau_y)$} (net.west);
  \draw[->, thick, netpurple]
        (net.north) to[bend right=20]
        node[pos=0.82,above,black]{$\widehat{w}$} (UPPER_P);
  \draw[->, thick, netpurple] (net.south) -- (law.north);
  \draw[->, thick, myorange]
        (law.west) to[bend left=16]
        node[pos=0.5,below,black]{$w_w$} (WALL_P);
\end{tikzpicture}%
}
\caption{Wall-only opposition control. The flow is an open channel at
$\Retau\!\approx\!180$ driven at constant flow rate; the lower wall carries
blowing and suction (orange out, teal in). Only the wall-shear components
$(\taux,\tauy)$ are sensed. A ConvGRU estimator $\mathcal{N}_\theta$ maps them to
the wall-normal velocity $\what$ on the detection plane $z^{+}\!\approx\!14$, and
the opposition law $w_w=-\alpha\,\what$ closes the loop at the wall.}
\label{fig:setup}
\end{figure}

Figure~\ref{fig:setup} shows the closed loop. At each control cycle the 
estimator reads the instantaneous wall-shear fields $(\taux,\tauy)$, predicts 
the detection-plane velocity $\what$, and feeds that prediction to the 
opposition law, which sets the wall actuation for the next cycle.

The estimator is a convolutional gated-recurrent network 
(ConvGRU~\cite{cho2014gru}): the wall input 
$\mathbf{s}_n=(\taux,\tauy)$ at cycle $n$ is compressed by a 
convolutional encoder into a feature field $\mathbf{x}_n$, a single 
ConvGRU cell updates a hidden state $\mathbf{h}_n$ that carries 
information from previous cycles, and a decoder reads out the 
prediction $\what_n$. 

A single wall snapshot under-determines the 
off-wall velocity field, since the wall signal constrains the large 
scales but carries little information about the small ones. Temporal 
context is useful because the recent history of the wall-shear signal 
encodes the advection and evolution of the near-wall structures, 
allowing the estimator to sharpen its prediction beyond what any 
instantaneous map could achieve. The recurrent update reads
\begin{align}
\mathbf{z}_n &= \sigma(\mathbf{W}_z \circledast 
    [\mathbf{x}_n,\mathbf{h}_{n-1}]),
\qquad
\mathbf{r}_n = \sigma(\mathbf{W}_r \circledast 
    [\mathbf{x}_n,\mathbf{h}_{n-1}]),
\nonumber\\
\tilde{\mathbf{h}}_n &= \tanh\!\big(\mathbf{W}_h \circledast
   [\mathbf{x}_n,\,\mathbf{r}_n\odot\mathbf{h}_{n-1}]\big),
\qquad
\mathbf{h}_n = (1-\mathbf{z}_n)\odot\mathbf{h}_{n-1}
             + \mathbf{z}_n\odot\tilde{\mathbf{h}}_n,
\label{eq:convgru}
\end{align}
with $\circledast$ a circular convolution, $\odot$ the Hadamard 
product, $\sigma$ the logistic sigmoid, and $[\,\cdot\,,\cdot\,]$ 
channel concatenation; $\mathbf{z}_n$ and $\mathbf{r}_n$ are the 
update and reset gates, which control how much of the previous hidden 
state is retained or overwritten at each cycle. All convolutions use 
$3\times3$ kernels with circular padding to respect the streamwise and 
spanwise periodicity of the channel, and the hidden state has $64$ 
channels. Table~\ref{tab:arch} lists the layers; the whole network has 
about $2.8\times10^{5}$ parameters.

The network is trained by backpropagation through time on sequences of 
twelve cycles, with the first three kept as recurrent warm-up and 
excluded from the loss. Three training stages are summarised in 
Table~\ref{tab:stages} and detailed in Appendix~\ref{app:training}: 
a first network is trained on converged opposition-control data, and 
two successive networks are each retrained on the closed-loop data 
generated by the previous one. The reason for this progressive 
retraining is developed in Section~\ref{sec:fix2}; it suffices here 
to note that the training distribution must match the flow state the 
controller actually visits, which the first network cannot guarantee.

\begin{table}[tbp]
\caption{The ConvGRU estimator. Every convolution uses a $3\times3$ kernel with
circular padding on the $256\times256$ plane, so the spatial size is preserved
throughout. The encoder lifts the two wall-shear channels to a feature volume,
one recurrent cell carries the hidden state $\mathbf{h}$ (64 channels) across
cycles, and the decoder reads out the prediction.}
\label{tab:arch}
\begin{tabular}{@{}llr@{}}
\toprule
block & operation (channels) & parameters \\
\midrule
input       & $(\taux,\tauy)$, $2$ channels                                 & -- \\
encoder     & conv $2\!\to\!32$, ReLU; conv $32\!\to\!64$, ReLU             & $19{,}104$ \\
ConvGRU cell& gate conv $128\!\to\!128$; candidate conv $128\!\to\!64$       & $221{,}376$ \\
decoder     & conv $64\!\to\!64$, ReLU; conv $1\!\times\!1$ $64\!\to\!1$     & $36{,}993$ \\
\midrule
total       &                                                               & $277{,}473$ \\
\botrule
\end{tabular}
\end{table}

\begin{table}[tbp]
\caption{The three training stages. Each trains the architecture of
Table~\ref{tab:arch} for thirty epochs on about one thousand control cycles
drawn from a different attractor (gen-0 instead uses ${\sim}5{,}000$ cycles of
OC data), with the per-channel normalisation set to that attractor's r.m.s. The
r.m.s.\ values track the attractor itself: the gen-0 controller inflates it,
and retraining brings it back toward the OC. The last column is the drag
reduction held over the $1000\,\tplus$ test window.}
\label{tab:stages}
\begin{tabular}{@{}llrrrr@{}}
\toprule
stage & training data & $\taux^{\mathrm{rms}}$ & $\tauy^{\mathrm{rms}}$
      & $w^{\mathrm{rms}}$ & DR (\%) \\
\midrule
gen-0     & OC attractor      & $1.05\!\times\!10^{-3}$ & $5.43\!\times\!10^{-4}$ & $8.85\!\times\!10^{-3}$ & $14.6$ \\
retrain-1 & gen-0 closed loop & $1.95\!\times\!10^{-3}$ & $7.66\!\times\!10^{-4}$ & $1.36\!\times\!10^{-2}$ & $21.0$ \\
retrain-2 & retrain-1 closed loop & $8.66\!\times\!10^{-4}$ & $5.34\!\times\!10^{-4}$ & $9.87\!\times\!10^{-3}$ & $19.7$ \\
\botrule
\end{tabular}
\end{table}

\subsection{Performance metrics}\label{sec:metrics}

Three quantities are monitored throughout. The drag
reduction compares the force required to drive the flow at fixed rate
against the uncontrolled baseline; since at constant flow rate that 
force is the mean streamwise body force,
\begin{equation}
\mathrm{DR} = 1 - \frac{\langle\Pi\rangle}{\langle\Pi_0\rangle},
\label{eq:dr}
\end{equation}
with $\Pi_0$ the uncontrolled value. The fidelity of the
wall-only estimate of the detection-plane velocity is measured by the
instantaneous spatial correlation between the predicted field $\what$ 
and the true wall-normal velocity $w(z_d)$,
\begin{equation}
\rho = \frac{\langle \what'\,w'\rangle}
            {\sqrt{\langle \what'^{2}\rangle\,\langle w'^{2}\rangle}},
\label{eq:corr}
\end{equation}
in which a prime denotes removal of the wall-parallel average; $\rho$ 
ranges from unity, when the estimator reproduces the detection-plane 
field exactly up to a scalar, to zero when the prediction is 
uncorrelated with the truth, and does so independently of whether the 
solver has yet diverged. The scale-by-scale counterpart of $\rho$ is 
the magnitude-squared coherence,
\begin{equation}
\gamma^{2}(\mathbf{k}) =
 \frac{\big|\langle \widehat{\what}(\mathbf{k})\,
 \widehat{w}^{*}(\mathbf{k})\rangle\big|^{2}}
      {\langle|\widehat{\what}(\mathbf{k})|^{2}\rangle\,
       \langle|\widehat{w}(\mathbf{k})|^{2}\rangle},
\label{eq:coh}
\end{equation}
where $\widehat{(\cdot)}$ is the wall-parallel Fourier transform, the 
average is over realisations, and $\gamma^{2}$ runs from unity 
(perfectly coherent) to zero (unrelated). The run length, defined as the number of control cycles completed
before the solution diverges, measures the numerical stability of the 
closed loop. It is important to note that run length and $\rho$ can move in opposite 
directions: a damped action delays divergence while the prediction 
decays, so run length alone is a misleading measure of controller 
performance.

\section{Results}\label{sec:results}

\subsection{Offline accuracy and closed-loop failure}\label{sec:paradox}

\begin{figure}[tbp]
\centering
\includegraphics[width=0.99\linewidth]{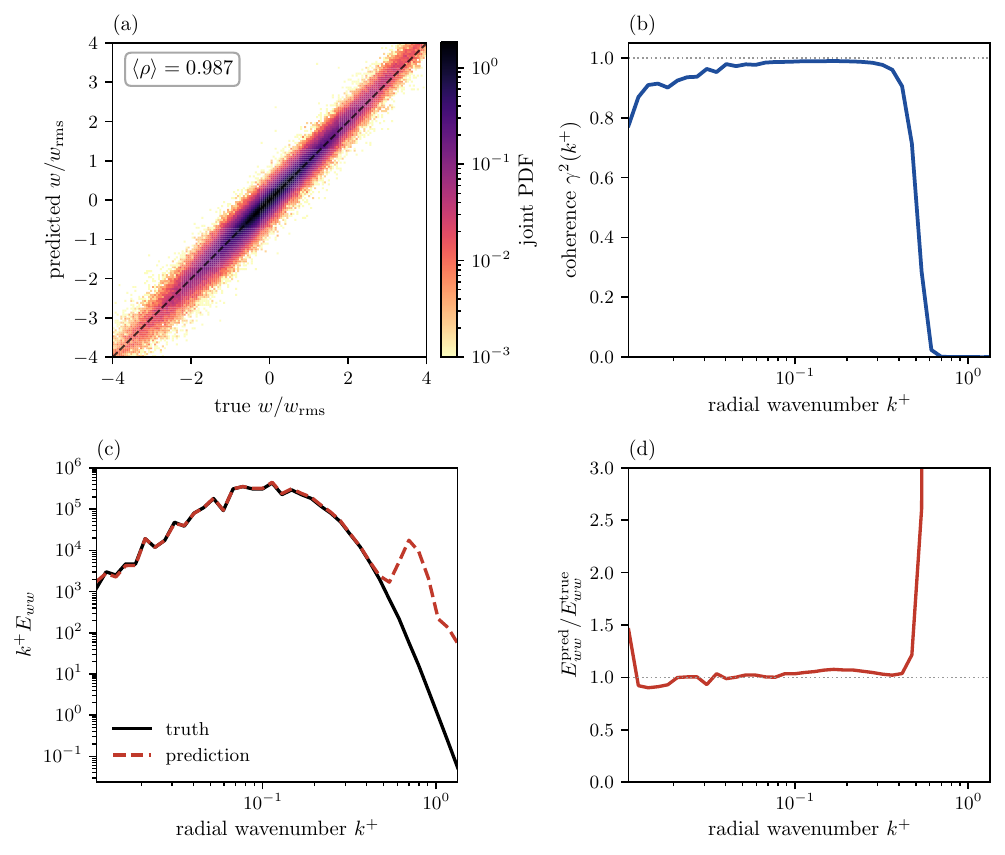}
\caption{Offline validation on OC data reserved for testing. (a) Joint
probability density of the predicted and true detection-plane velocity,
normalised by the root-mean-square of $w$; the dashed line is the 
diagonal and $\langle\rho\rangle$ the mean per-snapshot correlation. 
(b) Spectral coherence $\gamma^2(k)$ between prediction and truth, 
near unity through the energetic scales and collapsing only beyond 
$k^{+}\approx0.4$. (c) Premultiplied radial spectra $k\,E_{ww}(k)$ 
of the truth (black) and the prediction (red), which overlap through 
the energetic range. (d) Their ratio: close to unity where the flow 
is energetic, but rising well above it at high $k$, where the network 
injects a small \emph{excess} of energy that is energetically 
negligible offline.}
\label{fig:offline}
\end{figure}

Evaluated on OC data reserved for validation, the estimator 
reproduces the detection-plane velocity almost perfectly. 
Figure~\ref{fig:offline}(a) shows the joint distribution of predicted 
and true values collapsing onto the diagonal, with a mean spatial 
correlation of $0.987$ per snapshot; the validation loss tracks the training curve throughout, so the network is not overfitting. 
Figure~\ref{fig:offline}(b) shows the same agreement scale by 
scale: the prediction and the true field are fully coherent across the energy-containing range and lose coherence only beyond 
$k^{+}\approx0.4$, in the dissipative tail where the flow carries negligible energy.
The estimator predicts the detection-plane velocity from wall measurements with high accuracy, a result consistent with the offline reconstruction literature; whether that accuracy carries over to closed-loop control is a different question.

In particular, one feature of the offline results will prove critical in closed loop. Panels (c) and (d) of Figure~\ref{fig:offline} show that while the predicted spectrum overlaps the true one through the energetic range, it carries a small \emph{excess} of energy at the highest wavenumbers, where the coherence has collapsed and the network is in effect guessing. In the offline evaluation this excess is invisible: those scales hold so little energy that they do not affect the correlation or the loss by any measurable amount. 
In closed loop, this same energetically negligible excess has a disproportionate dynamical effect: once the estimator replaces the OC boundary condition, the correlation between the deployed field and the truth falls below one half within about five viscous times and reaches zero within fifteen to twenty, shortly after which the solution diverges. The transition is abrupt: there is no intermediate regime in which the wall-only controller delivers a degraded but useful fraction of the OC drag reduction.
%

\subsection{Mechanism of the instability}\label{sec:why}

Before analysing the neural estimator's failure, it is useful to 
establish how much of the detection-plane velocity the wall signal 
can in principle carry. The linear stochastic estimator (LSE) is the 
optimal linear map from wall measurements to the detection-plane 
velocity in the least-squares sense~\cite{suzuki2017estimation,
encinar2019wallview}. Its error at each wavenumber is set directly by 
the wall-to-plane coherence $\gamma^{2}$: where $\gamma^{2}$ is low, 
the wall signal does not contain that component of the detection-plane 
field, and no instantaneous map, however nonlinear, can recover what 
is not there.

Collecting the wall signal into $\mathbf{s}=(\taux,\tauy)$, the LSE 
is the mode-by-mode Wiener filter,
\begin{equation}
\widehat{\what}_{\rm LSE}(\mathbf{k}) =
   \mathbf{H}(\mathbf{k})\,\widehat{\mathbf{s}}(\mathbf{k}),
\qquad
\mathbf{H}(\mathbf{k}) =
   \langle \widehat{w}\,\widehat{\mathbf{s}}^{\dagger}\rangle\,
   \langle \widehat{\mathbf{s}}\,\widehat{\mathbf{s}}^{\dagger}
   \rangle^{-1},
\label{eq:lse}
\end{equation}
with $\dagger$ the conjugate transpose. Its residual is governed 
entirely by the wall-to-plane coherence~\eqref{eq:coh}: the fraction 
of detection-plane energy recoverable at wavenumber $\mathbf{k}$ by 
any linear map is $\gamma^{2}(\mathbf{k})$, so the reconstruction 
error cannot fall below $\langle|\widehat{w}|^{2}\rangle\,
[1-\gamma^{2}(\mathbf{k})]$. In the statistically steady 
opposition-controlled flow, $\gamma^{2}$ reaches about $0.85$ in the 
energy-containing band and falls toward $0.2$ at small scales; this 
is the upper bound on reconstruction accuracy against which the 
neural estimator must be assessed. The recurrent network of 
Section~\ref{sec:estimator} exceeds the instantaneous bound by 
drawing on the recent flow history, but it cannot recover information 
the wall does not carry at a given scale.

The first source of failure is that the controlled flow drifts away 
from the conditions on which the estimator was trained. The estimator 
was trained on data from the statistically steady flow under exact 
opposition control, a state the wall-only controller cannot maintain 
the instant its actuation differs from the OC's. 
This drift is rapid. The coherence between the wall stresses and the detection-plane 
velocity, about $0.85$ in the energetic band in the 
opposition-controlled flow, falls to roughly $0.45$ within a single control cycle under a sub-OC boundary condition.
The estimator is therefore asked, almost immediately, to predict flow states outside 
its training conditions, with no mechanism to detect or correct for 
this extrapolation. This observation explains why a single-snapshot estimator cannot succeed: without memory of the recent flow history, 
the estimator has no way to track this drift. A recurrent network 
mitigates this by carrying information across cycles, but as shown 
below, temporal memory alone is not sufficient to stabilise the loop.

\begin{figure}[tbp]
\centering
\includegraphics[width=\linewidth]{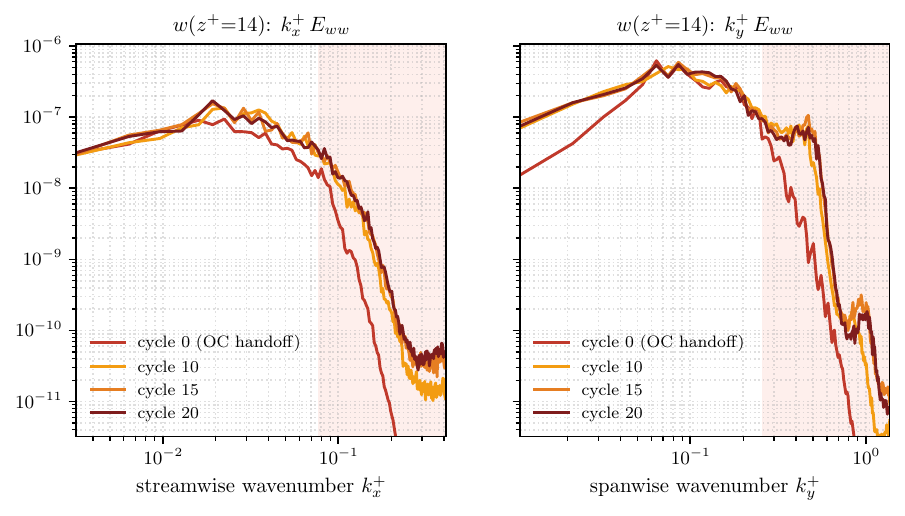}
\caption{The mechanism of divergence. Premultiplied spectra of the 
sampled detection-plane velocity at the moment of switching (red) 
and after $10$, $15$ and $20$ cycles (yellow to dark red) for the 
raw estimator with no spectral treatment. Energy piles up by about 
two decades in the high-wavenumber stop band (shaded) within ten 
cycles and then saturates; the energetic range is unchanged.}
\label{fig:divergence}
\end{figure}

What turns this drift into a rapid collapse is the way the flow 
responds to the high-wavenumber errors in the estimated wall-normal 
velocity. When this field is imposed as a wall boundary condition, 
the excess high-wavenumber energy it carries has a significant 
dynamical consequence. Figure~\ref{fig:divergence} shows the 
premultiplied spectrum of the wall-normal velocity at the detection 
plane through the first cycles after the estimator is switched on, 
without any spectral treatment. At the moment of switching the 
spectrum is well-behaved, with energy concentrated near 
$k^{+}\sim0.1$ and a steep roll-off at high wavenumbers. Within ten 
cycles the energy in the high-wavenumber band has grown by about two 
orders of magnitude, while the energy-containing range is essentially 
unchanged; the small-scale excess then saturates and persists. 
Viscous dissipation is insufficient to remove this injected 
high-wavenumber energy within a control cycle, so it accumulates 
and, through the nonlinear terms, contaminates the wall-stress fields 
the estimator reads.
The process forms a positive feedback loop: the corrupted wall-stress signal pushes the estimator inputs progressively further from its training distribution, degrading the prediction and injecting further small-scale energy at the wall boundary.

In control terms, a loop driven unstable by content it cannot model is
routinely tamed by attenuating that content before it closes the loop,
whether by limiting the amplitude or the bandwidth of the feedback
signal~\cite{skogestad2005multivariable}. Here the corruption builds up from
one cycle to the next, so the most natural attenuation is to lean on the
past: an exponential moving average (EMA) blends each new prediction with the
running history of earlier ones,
\begin{equation}
\bar{\what}_n = \beta\,\bar{\what}_{n-1} + (1-\beta)\,\what_n,
\qquad 0<\beta<1,
\label{eq:ema}
\end{equation}
where $\beta$ sets how much of that history is retained, so that the cleaner
estimates from before the corruption took hold dilute the degraded current
one. The same idea applies in space, by low-pass filtering the estimate
across the wall plane to remove the corrupted high wavenumbers, or in
amplitude, by lowering the gain $\alpha$ in~\eqref{eq:ocbc}. All three reduce
the corrupted signal reaching the wall and delay divergence, but smearing the
estimate in space or time causes the prediction to deteriorate with each
cycle.

\begin{figure}[tbp]
\centering
\includegraphics[width=0.7\linewidth]{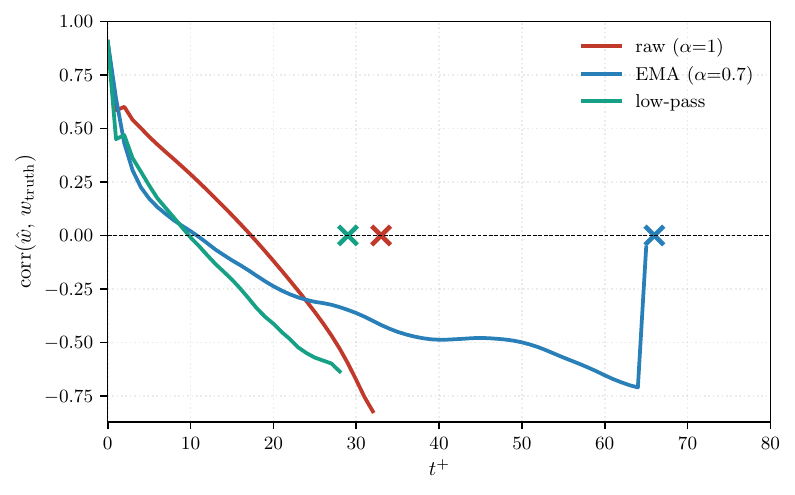}
\caption{Damping the action extends the run length but not prediction 
fidelity. Correlation between the estimated and true wall-normal 
velocity after the switch, for the raw estimator and two damped 
variants (EMA and spatial low-pass). Crosses mark the cycle at which 
each run diverges. The EMA survives about twice as long as the raw 
controller while holding a lower correlation throughout; all three 
fail well short of the $1000\,\tplus$ a working controller must 
reach.}
\label{fig:survival}
\end{figure}

Figure~\ref{fig:survival} confirms this by comparing the raw 
estimator with the two damped variants, using linear single-snapshot 
estimators to isolate the effect cleanly. The raw controller holds 
the best correlation in the early cycles yet diverges soonest, near 
$33\,\tplus$. The EMA roughly doubles the run length to about 
$66\,\tplus$, but at a lower correlation throughout. The low-pass 
behaves similarly. None of these reaches even a tenth of the 
$1000\,\tplus$ a working controller must sustain. Amplitude damping treats the symptom rather than the cause.

To address the problem, two remedies are presented in the following subsections: removing 
the high-wavenumber content the wall cannot inform before it reaches 
the wall, and retraining the estimator on the flow state the 
controller actually produces.
\subsection{Spectral matching of the deployed field}
\label{sec:fix1}

The first proposed remedy operates in the wavenumber plane on the 
predicted field, before it is imposed as a wall boundary condition. 
The exponential moving average introduced in Section~\ref{sec:why} 
is discarded: temporal blending preserves numerical stability at the 
cost of prediction accuracy. Figure~\ref{fig:survival} shows that this 
trade-off is not acceptable. Three operations are applied in sequence: 
a smooth spectral low-pass filter (see equation~\ref{eq:lowpass}); a 
spanwise spectral rescaling; and an element-wise amplitude cap retained 
only as a safeguard against rare excursions. The first two address the 
spectral diagnosis of Section~\ref{sec:why}; the third, as shown in 
Table~\ref{tab:abl}, is never active in practice.

\begin{figure}[tbp]
\centering
\includegraphics[width=\linewidth]{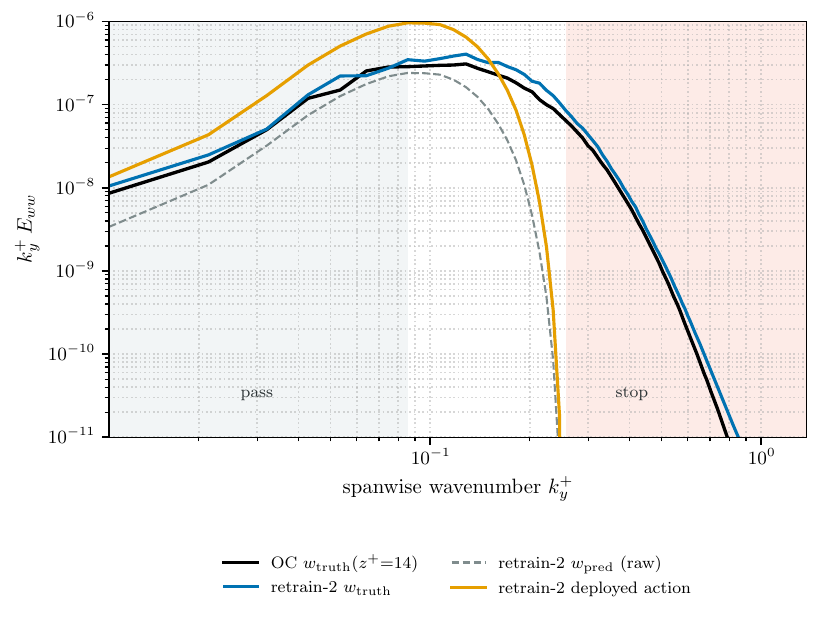}
\caption{Spectral matching in the spanwise direction. Premultiplied spanwise
spectrum of the detection-plane velocity, averaged over the steady portion of
the closed-loop run. Black: OC reference. Blue: retrain-2 true field. Grey
dashed: raw network output, with excess energy at high $k_y$. Orange: deployed
action after low-pass and spectral matching (rescaled to the detection-plane
level for comparison). Shaded bands mark the low-pass pass ($k_y^{+}\le0.085$)
and stop ($k_y^{+}\ge0.26$) regions.}
\label{fig:spectrum}
\end{figure}

The low-pass filter replaces a sharp spectral cut-off, which produces
the Gibbs phenomenon near the cut wavenumber, with a smooth raised-cosine
window applied separately in each wavenumber direction,
\begin{equation}
    \begin{aligned}
    \widehat{w}_b(k_x,k_y) &\mapsto G_x(k_x)\,G_y(k_y)\,
    \widehat{w}_b(k_x,k_y), \\[6pt]
    G_i(k) &=
    \begin{cases}
        1, & |k| \leq k_{\rm pass}^{i}, \\[4pt]
        \cos^2\!\left(\dfrac{\pi}{2}\,
        \dfrac{|k| - k_{\rm pass}^{i}}{k_{\rm stop}^{i} - k_{\rm pass}^{i}}
        \right), & k_{\rm pass}^{i} < |k| < k_{\rm stop}^{i}, \\[8pt]
        0, & |k| \geq k_{\rm stop}^{i}.
    \end{cases}
    \end{aligned}
    \label{eq:lowpass}
\end{equation}
for $i\in\{x,y\}$. Because the mesh is anisotropic
($\Delta x^{+}\!\neq\!\Delta y^{+}$), the corners are set independently in the
two directions but tied to a common physical wavelength, so the same cutoff
scale acts in the streamwise and spanwise directions.
These corners are not tuned for stability; the cutoff wavelength is read
directly off the energy spectrum of the training set, the same reference
field the estimator was taught to reproduce. The pass edge is placed at the
upper end of the energetic range of that spectrum, the band that carries
essentially all of the detection-plane wall-normal energy and over which the
offline prediction remains coherent with the truth (Figure~\ref{fig:offline}b).
The stop edge is placed where the reference spectrum has fallen to a
negligible fraction of its peak: beyond it the training flow contains no
appreciable energy, so any content the network deposits there cannot have
been informed by the wall and is by construction the spurious high-$k_y$
excess identified in Section~\ref{sec:why}. The corresponding spanwise bands
are shown against the reference spectrum in Figure~\ref{fig:spectrum}, with
the $\cos^2$ taper between them removing that excess without imposing a sharp
cut and the attendant Gibbs ringing.

The spectral matching is the more consequential of the two operations, 
motivated directly by the spectral diagnosis of Section~\ref{sec:why}. 
The spanwise spectral rescaling adjusts the predicted field so that its 
spanwise energy spectrum matches the reference spectrum 
$E_{\rm ref}(k_y)$ of the training set. This corrects the excess 
energy at high $k_y$ that the raw network output carries beyond the 
level observed in the statistically steady controlled flow, and which 
Section~\ref{sec:why} identified as the driver of the instability. 
Of the two spectral operations, the rescaling has the greater effect 
on closed-loop stability, as the results of Table~\ref{tab:abl} 
confirm. Each spanwise mode is rescaled according to
\begin{equation}
    \widehat{w}_b(k_x, k_y) \mapsto \widehat{w}_b(k_x, k_y)
    \sqrt{\frac{E_{\rm ref}(k_y)}{E_{w_b}(k_y)}}, \qquad
    E_{w_b}(k_y) = \sum_{k_x}
    \left|\widehat{w}_b(k_x, k_y)\right|^2.
    \label{eq:matching}
\end{equation}
The phase of each mode, which encodes the spatial organisation of the 
structures genuinely informed by the wall signal, is left unchanged; 
only the amplitude at each scale is constrained to match the reference 
statistics. The final element is an element-wise cap on the action, 
$\widehat{w}_b \mapsto \mathrm{clip}(\widehat{w}_b, -c_{\max}, 
c_{\max})$, retained only as a guard against rare excursions.

\begin{figure}[tbp]
\centering
\includegraphics[width=0.99\linewidth]{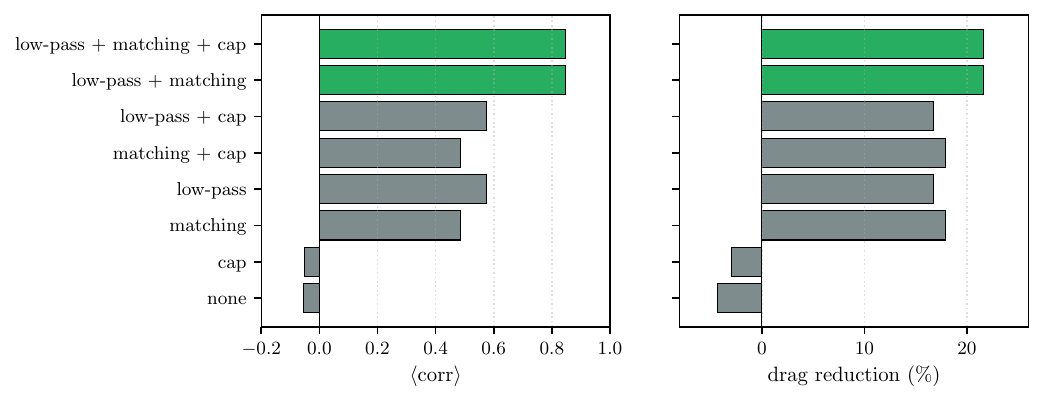}
\caption{Sensitivity of the deployed-field operations over $200\,t^+$ 
from the OC restart with the gen-0 estimator. Green bars mark 
configurations with both the low-pass and the spanwise spectral 
matching active. Left: mean correlation between the deployed action 
and the true detection-plane velocity. Right: mean drag reduction.}
\label{fig:ablation}
\end{figure}

\begin{table}[tbp]
\caption{Effect of the individual deployed-field operations over 
$200\,t^+$ from the OC restart with the gen-0 estimator, assessed 
by activating each operation in isolation and in combination. 
$\langle\rho\rangle$ is the mean correlation between the deployed 
action and the true detection-plane velocity, DR the drag reduction, 
$a_{\max}$ the cycle-mean of the largest action magnitude (the cap is 
$0.128$), and $\langle r \rangle$ the root-mean-square ratio of 
predicted to true field.} \label{tab:abl}
\begin{tabular}{@{}lrrrr@{}}
\toprule
configuration & $\langle\rho\rangle$ & DR (\%) & $a_{\max}$ & $\langle r\rangle$ \\
\midrule
low-pass $+$ matching $+$ cap & 0.85 & 21.6 & 0.038 & 0.88 \\
low-pass $+$ matching         & 0.85 & 21.6 & 0.038 & 0.88 \\
low-pass $+$ cap              & 0.58 & 16.8 & 0.119 & 1.09 \\
matching $+$ cap              & 0.49 & 17.9 & 0.052 & 0.85 \\
low-pass only                 & 0.58 & 16.8 & 0.121 & 1.09 \\
matching only                 & 0.49 & 17.9 & 0.052 & 0.85 \\
cap only                   & $-0.05$ & $-2.9$ & 0.128 & 4.16 \\
none                       & $-0.05$ & $-4.3$ & 0.162 & 4.05 \\
\botrule
\end{tabular}
\end{table}
Table~\ref{tab:abl} and Figure~\ref{fig:ablation} show the 
contribution of each operation individually and in combination, 
over a two-hundred-cycle window from the OC restart.
Both the low-pass filter and the spectral matching improve the correlation 
and the drag reduction independently, and their effects are nearly additive.
It is noted that either operation alone lifts the correlation $\rho$ between the 
deployed field and the true detection-plane velocity from essentially 
zero to between $0.49$ and $0.58$, and recovers seventeen to eighteen 
percent drag reduction.
Together they reach 
a mean correlation of $0.85$ and a drag reduction of $21.6\%$, 
recovering the OC level. The amplitude cap contributes nothing 
to these gains: in the working configuration the largest action 
magnitude averages $0.038$, well below the cap of $0.128$, and 
removing the cap leaves the results unchanged. Moreover, when both spectral 
operations are switched off, the cap alone cannot keep the correlation 
positive. 

Figure~\ref{fig:spectrum} shows the effect of the matching in the 
spanwise spectrum, which the diagnosis of Section~\ref{sec:why} 
identified as the critical variable. The raw network output (grey) 
carries the excess spanwise energy at high $k_y$ that drives the 
instability. The low-pass filter and the spectral rescaling together 
remove that excess and bring the surviving band into agreement with 
the reference envelope, so that the deployed action (orange) 
coincides with the true detection-plane field (black) through the 
energetic range. The residual disagreement is confined to the 
spanwise roll-off near $k_y^+ \sim 0.09$--$0.17$, where the 
raised-cosine taper remains active, and amounts to at most a factor 
of two in spectral energy. With this treatment in place the closed 
loop no longer diverges; the controller remains stable indefinitely. 
Over long times, however, the drag reduction falls short of the OC
level, for the same reason as the original failure: the estimator 
operates on a flow state that differs from the one on which it was 
trained, as discussed in Section~\ref{sec:fix2}.
%
%

\subsection{Closed-loop retraining}\label{sec:fix2}
\begin{figure}[tbp]
\centering
\includegraphics[width=0.74\linewidth]{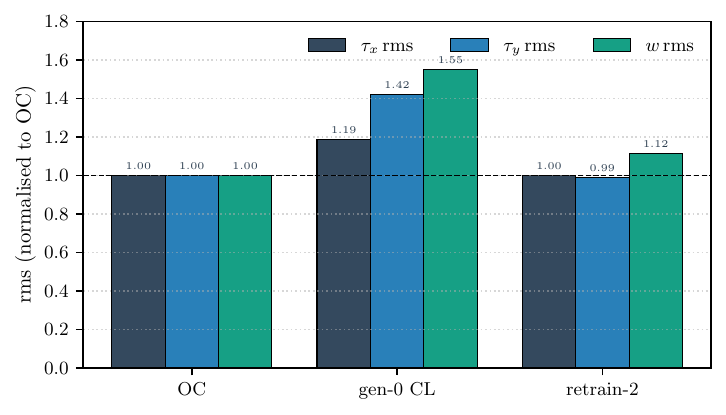}
\caption{Retraining realigns the closed-loop attractor with the OC.
Root-mean-square statistics on the closed-loop trajectory, normalised by their
OC values, for the gen-0 estimator (which sits on an inflated attractor) and
after one and two rounds of closed-loop retraining. The wall-shear components
return to within $\sim1\%$ of OC; the wall-normal velocity retains a
$\sim11\%$ residual.}
\label{fig:attractor}
\end{figure}

Spectral matching stabilises the closed loop but leaves a controller 
that operates on flow states outside the OC training set. The stabilised gen-0 
estimator, trained on OC data, remains stable over the full thousand-cycle 
test window.  However, it delivers only $14.6\%$ drag reduction, well short 
of the standard OC's $22.7\%$. 
The cause is the same distribution mismatch that drove the original failure,
now bounded rather than explosive. Spectral matching suppresses the
small-scale instability but does not realign the controlled flow with
the OC training set: the estimator receives wall measurements drawn from a
statistical state it was never trained on.
The distance between the OC statistical state and that of the
estimator trained on OC data (gen-0) is quantified by the
root-mean-square statistics of the wall-shear and wall-normal
velocity fields.
Figure~\ref{fig:attractor} shows 
that on the gen-0 closed-loop trajectory the r.m.s.\ values of 
$\tau_x$, $\tau_y$ and $w$ exceeds the OC values by twenty to fifty percent:
the controller has not diverged, but the two flows develop into different statistical states.
The remedy is to retrain the estimator on data drawn from the
closed-loop flow itself. At each cycle of the gen-0 closed-loop
run the simulation records the wall quantities and the
corresponding detection-plane velocity; these input--output
pairs form the new training set, with the wall inputs
normalised by their r.m.s. over this run rather than by the
OC values. The network is then refitted on this set, with the
detection-plane velocity as target, at no additional
computational cost since the simulation supplies it at every
cycle regardless.

This iterative approach rests on a simple principle: each estimator
is trained on data from the flow state it will actually control,
rather than from the reference OC state. The same idea underlies
iterative data-collection schemes in learning-based
control~\cite{ross2011dagger} and neural-network corrections to
numerical \textsc{pde} solvers~\cite{um2020solverloop}.

One aspect of the closed-loop retraining procedure differs from
standard offline fitting: the actuation modifies the flow from
which subsequent training data are drawn, so each iteration
operates on a flow shaped by the previous estimator, and the
procedure continues until the change in actuation between
iterations falls below a prescribed threshold. Since the
\textsc{DNS} records the wall-sensor measurements and the true
detection-plane velocity at every control cycle, the training
data are already available as a byproduct of the closed-loop
run; no adjoint computation or separate data-collection campaign
is needed. This makes the procedure attractive for experimental settings,
where only wall-sensor measurements are available.

%
Deployed under the spectral rescaling, retrain-1 initially
recovers a drag reduction of approximately $21\%$ but drifts
downward over the window, indicating that one iteration does not
suffice. A second iteration, retrain-2, trained on the
closed-loop data produced by retrain-1, stabilises at a mean of
$19.7\%$ over the full thousand-cycle window, a figure which is only marginally below  the
standard OC performance. Figure~\ref{fig:attractor}
shows that the root-mean-square wall-shear statistics are within
approximately one percent of the OC values, with a residual
discrepancy of approximately eleven percent in the wall-normal
velocity, addressed in Section~\ref{tito}.


\begin{figure}[tbp]
\centering
\includegraphics[width=\linewidth]{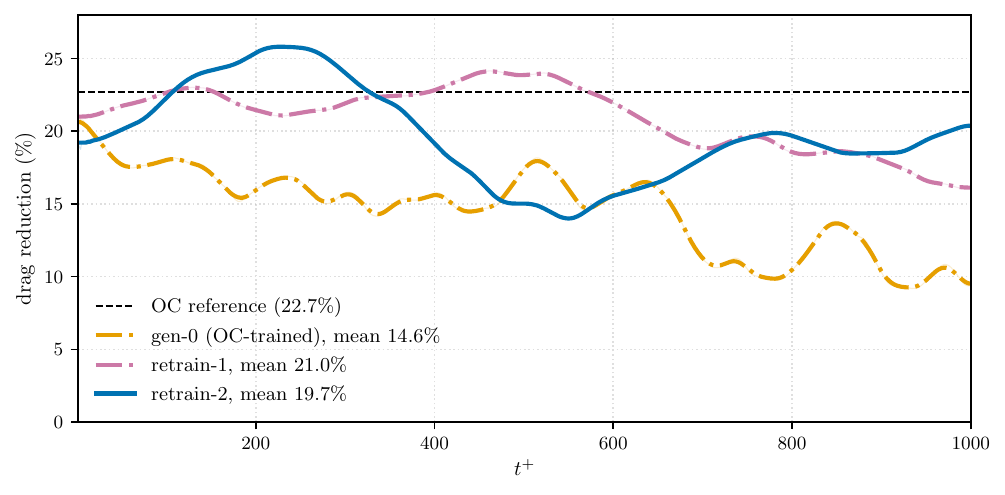}
\caption{Drag reduction over the closed-loop window for the three types of
estimator under the same spectral treatment. The OC-trained controller
(gen-0) drifts off the training attractor; one retraining iteration recovers only initially
near-OC drag reduction, while a second produces consistent values. Lines are
$25$-cycle running means over the per-cycle traces (faint); legend values are
full-window means.}
\label{fig:dr}
\end{figure}

Figure~\ref{fig:dr} reports the drag reduction over the
thousand-cycle window for all three estimators, deployed
under the same spectral treatment. The gen-0 controller starts
near the OC level and drifts steadily downward as the controlled
flow departs from the statistical state on which it was trained.
Retrain-1 recovers to approximately $21\%$ initially but exhibits
a similar, although less severe, drift. Retrain-2 eliminates the drift
and holds a stable mean of $19.7\%$ across the full window.
The window spans approximately $5.4$ large-eddy turnover times,
long enough for any systematic drift to have appeared. A stable wall-only controller at near-OC
performance is therefore achievable once the spectral constraint
is in place and the estimator has been retrained on closed-loop
data: the former keeps the deployed field consistent with the
statistics of the controlled flow, and the latter aligns the
estimator's training data with the flow it operates on.

\subsection{The controlled flow}\label{sec:flow}
A controller may match the correct mean drag while sustaining a
qualitatively different near-wall turbulence structure. To address
this, both OC and the twice-retrained wall-only controller were
run for a further thousand viscous time units. Wall-parallel-averaged
profiles of the mean velocity, velocity variances, and Reynolds
shear stress were accumulated, together with one- and two-point
statistics of the detection-plane fields; this comparison follows
the approach of several studies of data-driven wall
controllers~\cite{hammond1998oppositionmech,varela2022actuators,zhou2025drlhighre}.

Recovering the drag reduction confirms that the wall-only controller
reproduces one integral quantity of the OC state, but leaves open the
more demanding question of whether it reproduces the flow itself.
A controller may reproduce the correct mean drag while sustaining a
qualitatively different near-wall turbulence structure.
To address this, both OC and the twice-retrained wall-only controller
were run for a further thousand viscous time units with complete
statistical sampling.
Wall-parallel-averaged profiles of the mean velocity, velocity
variances, and Reynolds shear stress were accumulated, together
with one- and two-point statistics of the detection-plane fields,
for comparison with previous wall-control
studies~\cite{hammond1998oppositionmech,varela2022actuators,zhou2025drlhighre}.

Throughout this section, statistics are expressed in inner units
based on each flow's mean friction velocity, averaged over the
statistical sampling window; the measured values are
$u_\tau = 0.0560$ for OC and $u_\tau = 0.0574$ for the
wall-only controller, corresponding to friction Reynolds numbers
of $Re_\tau = 161$ and $Re_\tau = 165$, respectively.
The wall-only controller operates at a marginally higher near-wall
energy level, consistent with the drag gap quantified in
Section~\ref{sec:fix2}.

\begin{figure}[tbp]
\centering
\includegraphics[width=\linewidth]{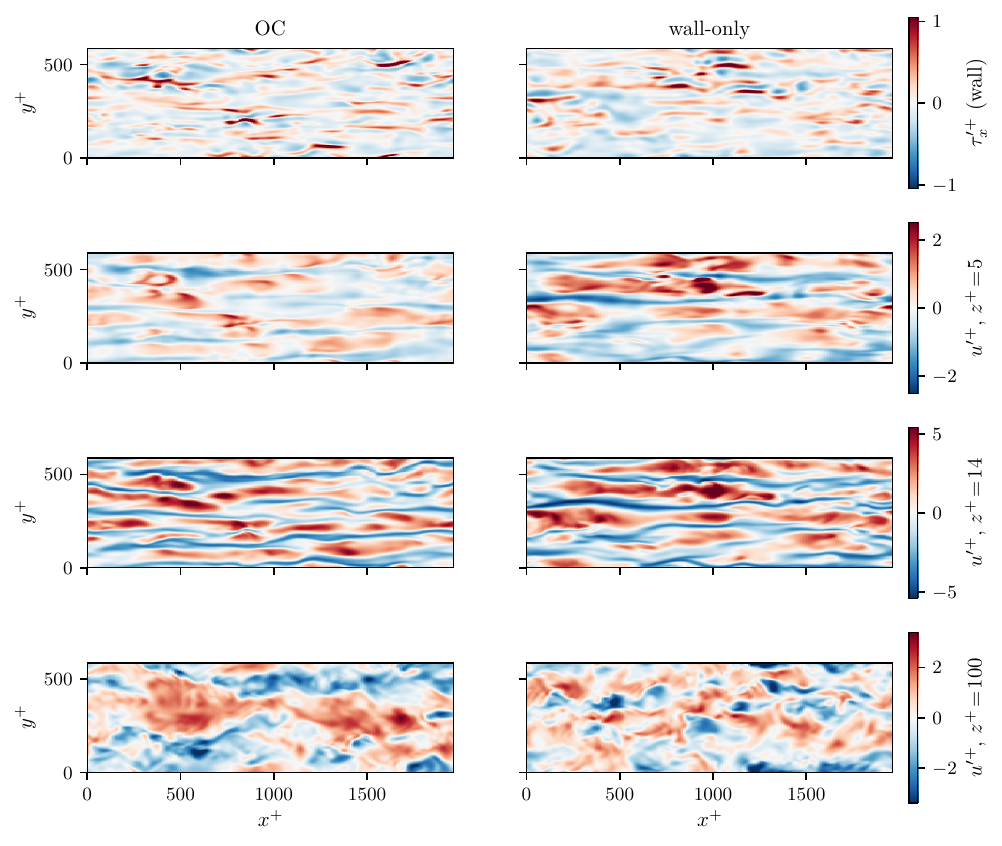}
\caption{Instantaneous streamwise field in wall-parallel planes, OC (left)
versus the wall-only controller (right). From top: wall-shear
fluctuation $\tau_x'^{+}$, then the streamwise fluctuation $u'^{+}$
at $z^{+}\!=\!5$, $14$, and $100$ (plane indices read from the
stretched-grid $z$-coordinate); colour scales are shared within each
row so that amplitudes are directly comparable between the two flows.}
\label{fig:snap_u}
\end{figure}

\begin{figure}[tbp]
\centering
\includegraphics[width=\linewidth]{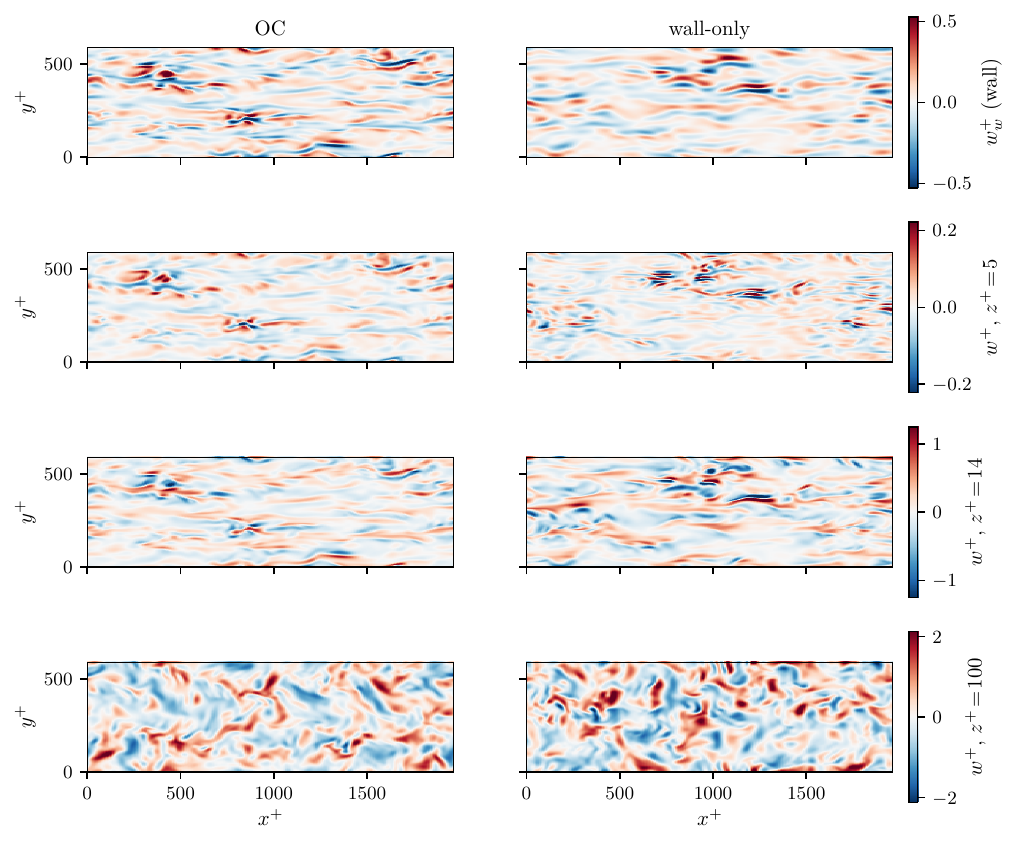}
\caption{As figure~\ref{fig:snap_u}, but for the wall-normal field.
From top: the wall actuation $w_w^{+}$, then
$w^{+}$ at $z^{+}\!=\!5$, $14$ and $100$. Each row shares a colour scale.}
\label{fig:snap_w}
\end{figure}

The most direct comparison is visual inspection of instantaneous
snapshots. Figures~\ref{fig:snap_u} and~\ref{fig:snap_w} show
wall-parallel planes of the streamwise and wall-normal velocity
fields at four wall-normal distances ($z^{+}\!=\!0$, the wall;
$z^{+}\!=\!5$; $14$; and $100$). At the wall the relevant
quantities are the streamwise wall-shear fluctuation $\tau_x'^{+}$
and the imposed actuation $w_w^{+}$. Above the wall, the
streamwise fluctuation $u'^{+}$ is organised into the elongated
low- and high-speed streaks of the near-wall cycle, and the
wall-normal velocity $w^{+}$ fills the remaining panels.

The OC and wall-only cases are visually indistinguishable at
every wall distance: streak spacing, streamwise coherence, and the
small-scale structure of the wall-normal field are the same in
both. This is a first indication that the wall-only controller has
reproduced the near-wall turbulence structure, not merely the
integrated drag.

Figure~\ref{fig:snap_w} also reveals a difference that is
expected. In classical OC the wall actuation $w_w^{+}$ is, by
construction, the negative of the detection-plane velocity it
opposes, so the wall and $z^{+}\!=\!14$ panels are mirror images.
For the wall-only controller, the actuation is the negative of the
\emph{estimated} detection-plane velocity rather than the true
field; the wall-normal velocity at the wall and at $z^{+}\!=\!14$
are therefore clearly anti-correlated but no longer exact mirror
images. The wall-only controller thus imposes a wall boundary
condition that differs slightly from the OC reference, as is
inherent in the use of an imperfect estimator. The rest of this
section makes the comparison quantitative.

\begin{figure}[tbp]
\centering
\includegraphics[width=\linewidth]{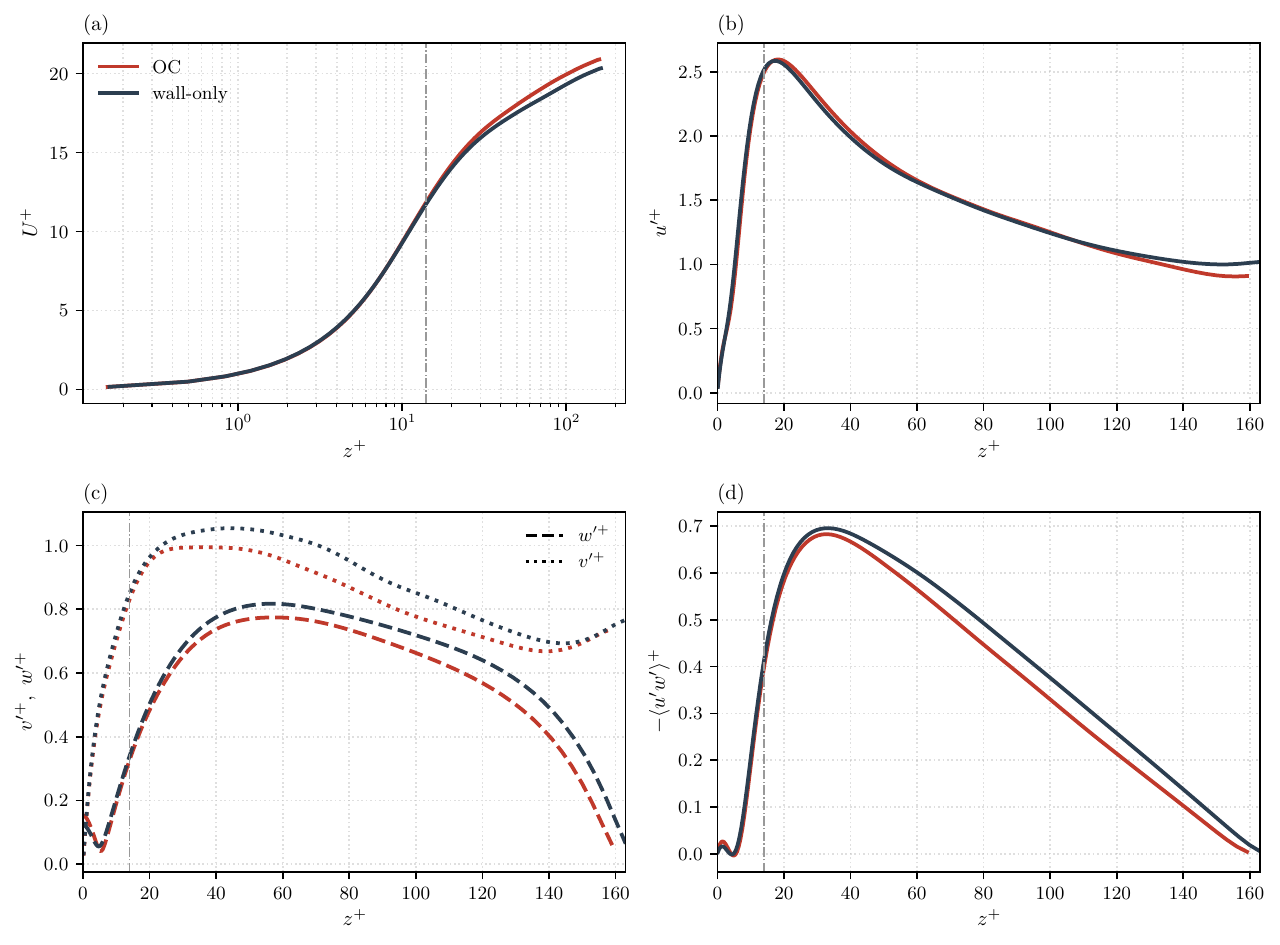}
\caption{Channel statistics of OC (red) and the wall-only controller case (dark), each scaled in its own wall units and plotted against $z^{+}$. (a) Mean streamwise velocity
$U^{+}$; (b) streamwise r.m.s.\ $u'^{+}$; (c) wall-normal $w'^{+}$ (dashed) and
spanwise $v'^{+}$ (dotted) r.m.s.; (d) Reynolds shear stress
$-\langle u'w'\rangle^{+}$. The dot-dashed line marks the detection plane
$z^{+}\!\approx\!14$.}
\label{fig:profiles}
\end{figure}

When the mean velocity profiles are compared
(Figure~\ref{fig:profiles}a), they collapse across the entire
channel height. The streamwise fluctuation (panel b) is
reproduced, with the wall-only controller marginally stronger near
the buffer-layer peak. The wall-normal and spanwise fluctuations
(panel c) agree closely, and the Reynolds shear stress (panel d)
is reproduced in both magnitude and shape, peaking near
$z^{+}\!\approx\!30$, with the wall-only controller marginally
higher in the outer region. 
These small discrepancies are quantitative and consistent with
the slightly lower drag reduction of the wall-only controller. These one-point statistics indicate that the near-wall turbulence
structure is almost unchanged.

%
%
%

\begin{figure}[tbp]
\centering
\includegraphics[width=0.92\linewidth]{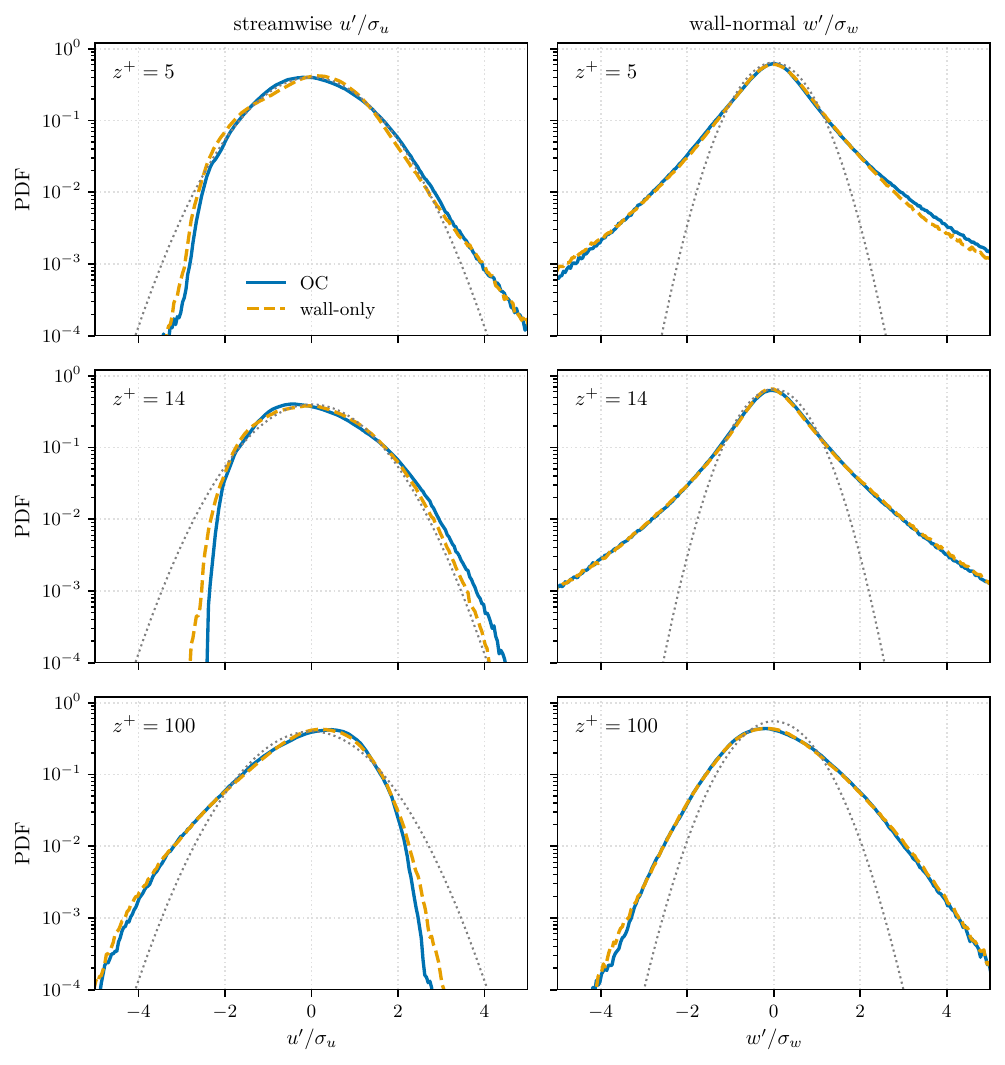}
\caption{Probability density functions of the standardised streamwise (left) and
wall-normal (right) fluctuations at $z^{+}=5$, $14$ and $100$ (rows), OC (blue,
solid) versus the wall-only policy (orange, dashed). Dotted curves: the standard
unit Gaussian for $u'$, and a Gaussian fitted to the core ($|s|<1.5$) for the
intermittent $w'$. The wall-normal fluctuation is strongly non-Gaussian near the
wall and relaxes towards Gaussian by $z^{+}=100$.}
\label{fig:z14pdf}
\end{figure}

The agreement extends beyond second-order moments to the full
velocity distributions. Figure~\ref{fig:z14pdf} compares
probability density functions of the streamwise and wall-normal
fluctuations at $z^{+}=5$, $14$ and $100$. The expected
wall-normal trends appear in both flows: the streamwise
fluctuation is positively skewed and sub-Gaussian (lighter-tailed
than a normal distribution) close to the wall, turning negatively
skewed in the outer region ($z^{+}=100$); the wall-normal
fluctuation is strongly intermittent near the wall, with flatness
of order eight to ten at $z^{+}=5$ and $14$, relaxing towards
Gaussian by $z^{+}=100$ (flatness $\approx3.6$). At each wall
distance, OC and the wall-only controllers superpose across three
decades of probability, with only marginal differences in the
distribution tails, confirming that the wall-only controller
reproduces the near-wall intermittency of the controlled flow,
not merely its second-order moments.

%

\begin{figure}[tbp]
\centering
\includegraphics[width=0.78\linewidth]{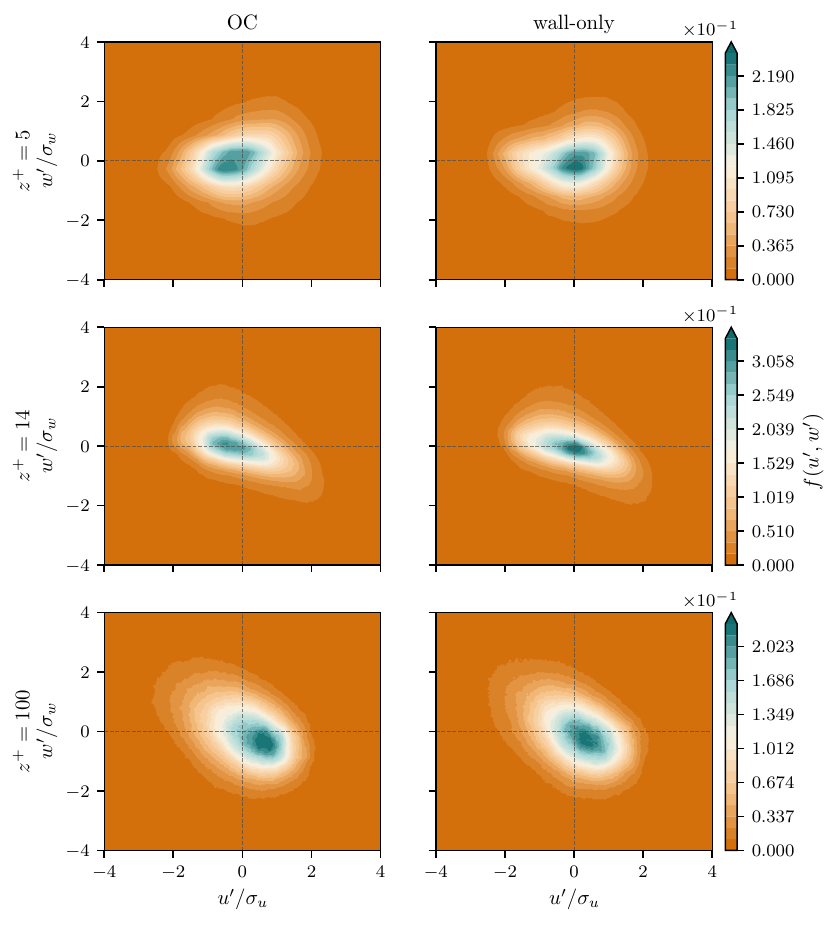}
\caption{Joint PDF $f(u',w')$ of the streamwise and wall-normal fluctuations at
$z^{+}=5$, $14$ and $100$ (rows), OC (left) versus the wall-only policy (right).
Each row uses its own colour scale, shared between the two flows. Quadrant
occupancies at the detection plane are collected in Table~\ref{tab:quad}.}
\label{fig:z14quad}
\end{figure}

\begin{figure}[tbp]
\centering
\includegraphics[width=0.78\linewidth]{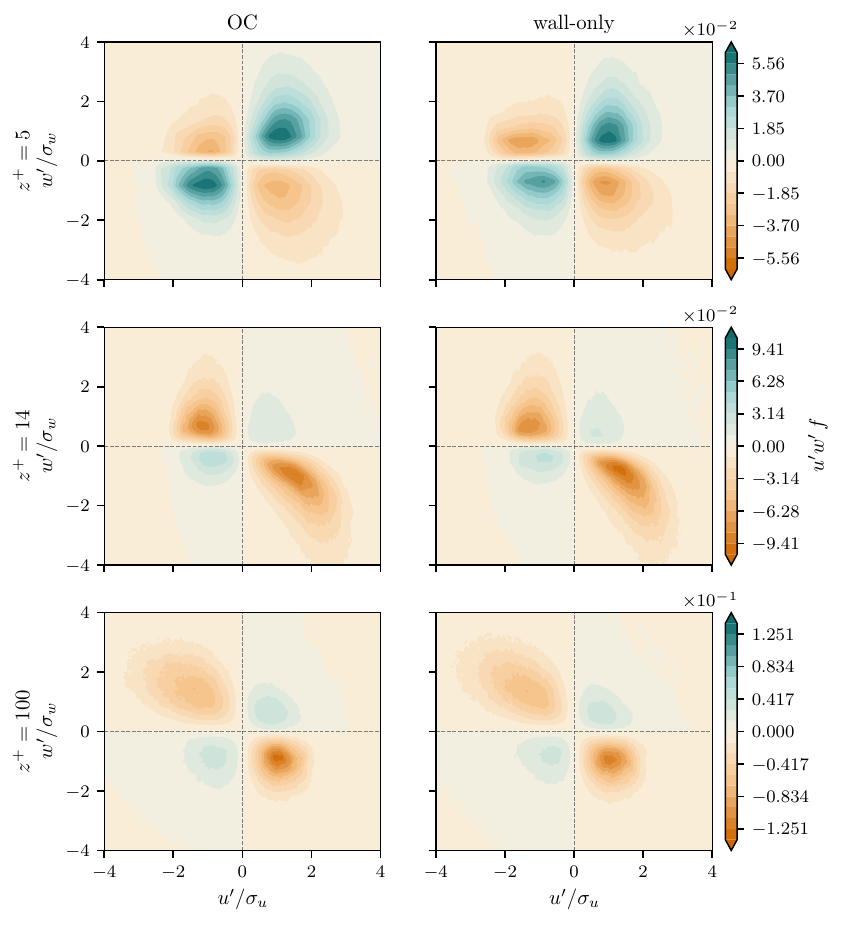}
\caption{Reynolds-shear-weighted joint density $u'w'\,f(u',w')$ at $z^{+}=5$, $14$
and $100$ (rows), OC (left) versus the wall-only policy (right). The Q2 (ejection)
and Q4 (sweep) lobes carry the momentum flux; the interaction quadrants Q1 and Q3
contribute the opposite sign. Each row uses its own colour scale, shared between
the two flows.}
\label{fig:z14premult}
\end{figure}

Since skin-friction drag is set by the near-wall Reynolds shear
stress~\cite{fukagata2002fik}, a sharper test uses the joint
statistics of $u'$ and $w'$. The joint probability density
(Figure~\ref{fig:z14quad}, OC and wall-only controller,  side by side) is
tilted into the second and fourth quadrants at every height, the
tilt sharpening with distance from the wall. Its stress-weighted
form $u'w'\,f(u',w')$ (Figure~\ref{fig:z14premult}) has two
lobes that generate the momentum flux: ejections of low-speed
fluid away from the wall (Q2: $u'<0$, $w'>0$) and sweeps of
high-speed fluid towards it (Q4: $u'>0$, $w'<0$).

The complementary quadrants Q1 and Q3, which contribute the
opposite sign to the Reynolds shear stress~\cite{wallace1972wallregion,
lu1973reynoldsstress,lozanoduran2012momentum}, account for far
fewer events: in OC they represent only $14$--$19\%$ of the
total against approximately $35\%$ and $33\%$ in Q2 and Q4
(Table~\ref{tab:quad}). The wall-only controller reproduces both the quadrant event
fractions and the stress distribution almost exactly. 
Each row in
Figures~\ref{fig:z14quad} and~\ref{fig:z14premult} uses a colour
scale shared between the two flows, so that peak height and shape
are directly comparable; the two cases are nearly indistinguishable
in every panel.

%
%

\begin{table}[tbp]
\caption{Quadrant occupancy at $z^{+}\!\approx\!14$: the percentage of samples
falling in each quadrant (Q1 outward, Q2 ejection, Q3 inward, Q4 sweep). The
ejection (Q2) and sweep (Q4) quadrants are the most populated in both flows, and
the policy reproduces the OC occupancy closely; the four values sum to $100\%$.}
\label{tab:quad}
\begin{tabular}{@{}lrrrr@{}}
\toprule
 & Q1 (outward) & Q2 (ejection) & Q3 (inward) & Q4 (sweep) \\
\midrule
OC        & $14.1$ & $34.7$ & $19.2$ & $32.0$ \\
wall-only & $14.5$ & $33.5$ & $18.1$ & $33.9$ \\
\botrule
\end{tabular}
\end{table}

\begin{figure}[tbp]
\centering
\includegraphics[width=0.82\linewidth]{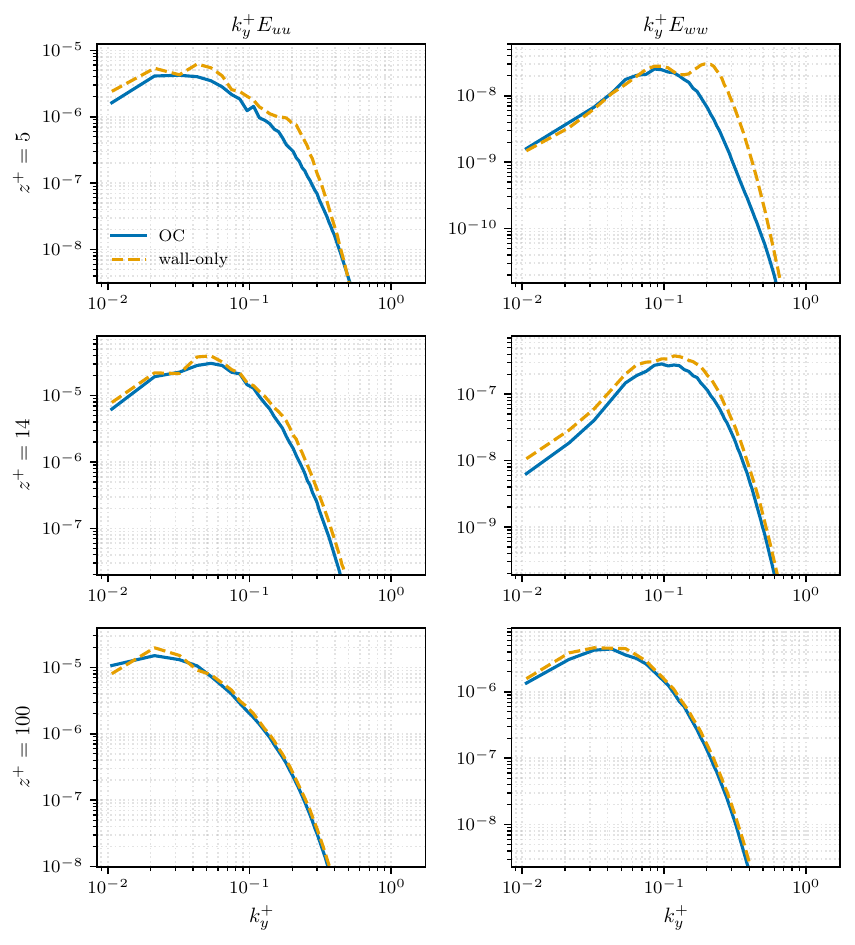}
\caption{Premultiplied spanwise energy spectra at $z^{+}=5$, $14$ and $100$
(rows), OC (blue, solid) versus the wall-only policy (orange, dashed): streamwise
energy $k_y^{+}E_{uu}$ (left) and wall-normal energy $k_y^{+}E_{ww}$ (right).}
\label{fig:z14spec}
\end{figure}

Finally, the spanwise energy spectra of the wall-parallel fields
(Figure~\ref{fig:z14spec}) confirm that the agreement holds scale
by scale. At all three heights the streamwise and wall-normal
energy spectra collapse onto the OC reference across the spanwise
band; the wall-normal energy peaks near $k_y^{+}\!\sim\!0.1$ at
the detection plane and shifts to larger spanwise scales away from
the wall. The only systematic departure is a slightly raised
wall-normal tail at high wavenumber, the fingerprint of the
heavier $w'$ tails and the small high-wavenumber surplus that the
spectral matching of Section~\ref{sec:fix1} bounds but does not
entirely remove. Taken together, the mean profiles, one-point
statistics, quadrant event fractions, and energy spectra show that
the wall-only controller reproduces the OC turbulence structure,
differing only by the slight excess of near-wall activity expected
of a marginally weaker controller.

\begin{figure}[tbp]
\centering
\includegraphics[width=0.78\linewidth]{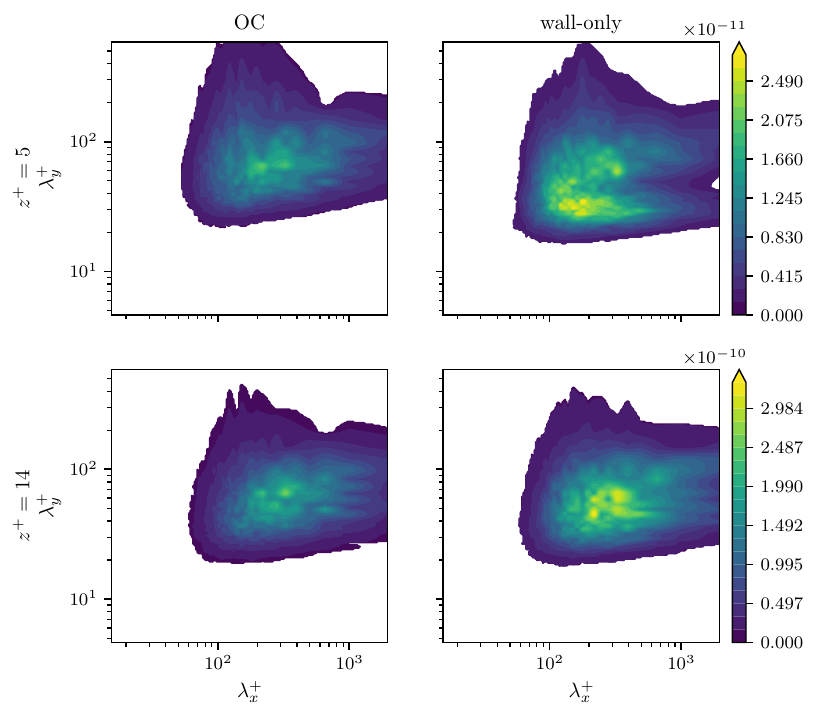}
\caption{Two-dimensional premultiplied wall-normal energy spectrum
$k_xk_y\Phi_{ww}$ at $z^{+}=5$ and $14$ (rows) in the
$(\lambda_x^{+},\lambda_y^{+})$ plane, OC (left) versus the wall-only policy
(right). Each row shares a colour scale between the two flows and cells below the
top $95\%$ of the energy are blank. Viscous units use the OC friction velocity.}
\label{fig:z142dww}
\end{figure}

\begin{figure}[tbp]
\centering
\includegraphics[width=0.78\linewidth]{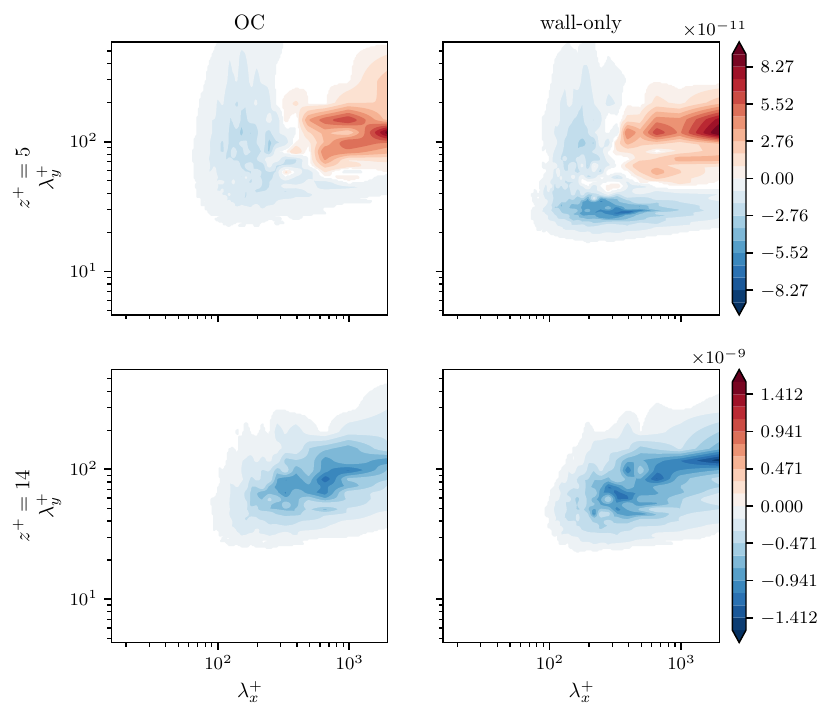}
\caption{Two-dimensional premultiplied $u'w'$ cospectrum
$k_xk_y\,\mathrm{Co}_{uw}$ at $z^{+}=5$ and $14$ (rows) in the
$(\lambda_x^{+},\lambda_y^{+})$ plane, OC (left) versus the wall-only policy
(right). Blue marks the scales with $u'w'<0$ that carry the Reynolds shear stress
down the mean gradient, and whose integral is the mean stress. Red marks scales
with $u'w'>0$, a counter-gradient (outward and inward) momentum flux that acts
against the mean shear; here it is confined to the largest streamwise scales and
carries little of the total. Each row shares a colour scale between the two flows,
and cells below the top $95\%$ of the magnitude are blank. Viscous units use the
OC friction velocity.}
\label{fig:z142dco}
\end{figure}

The same comparison in two dimensions reveals where in the
$(\lambda_x^{+},\lambda_y^{+})$ plane (streamwise and spanwise
wavelengths, respectively) the wall-normal energy and the shear
stress reside, at the near-wall and detection planes
($z^{+}=5$ and $14$). The wall-normal energy $\Phi_{ww}$
(Figure~\ref{fig:z142dww}) and the $u'w'$ cospectrum
(Figure~\ref{fig:z142dco}) concentrate at the compact scales
$\lambda_x^{+}\!\sim\!200$, $\lambda_y^{+}\!\sim\!100$ that
carry most of the Reynolds shear stress, the structures coarsening
with distance from the wall. OC and the wall-only controller
show these features at the same scales at both heights; the
wall-only controller is slightly more energetic throughout, most
visibly in the cospectrum, once more the signature of marginally
weaker control.


\subsection{Residual gap to opposition control} \label{tito}

The twice-retrained controller closely approaches the OC case but does
not perfectly match it. Quantifying the gap between the two cases allows to establish what wall-only
sensing can achieve and where further improvement might be sought.
The drag reduction falls approximately three percentage points
short of the OC, and the correlation between the wall actuation and
the detection-plane velocity it opposes is $0.80$, against $0.95$
for OC. This residual prediction error is consistent with the wall-to-detection-plane coherence ceiling established in
Section~\ref{sec:why}, and the controller is correspondingly
slightly weaker than the OC one.

\begin{figure}[tbp]
\centering
\includegraphics[width=0.99\linewidth]{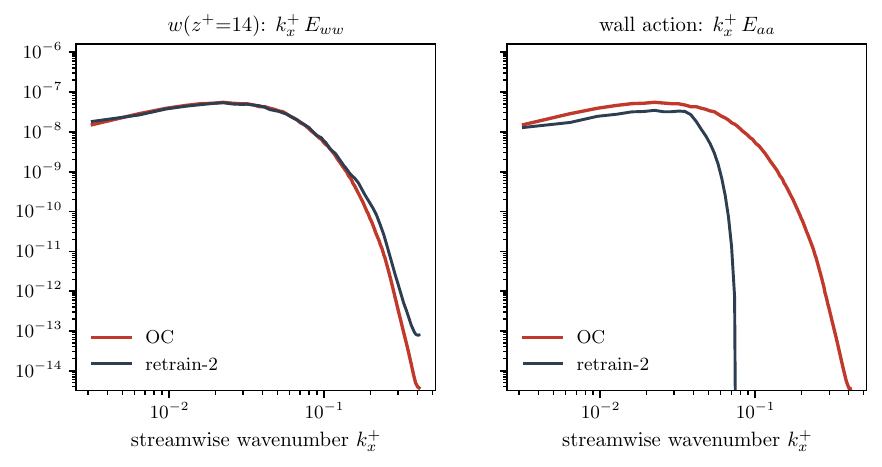}
\caption{Streamwise premultiplied spectra of the detection-plane velocity (left)
and of the deployed action (right), averaged over the steady portion of each run.
Red: OC. Dark blue: twice-retrained controller.}
\label{fig:gap_kx}
\end{figure}

\begin{figure}[tbp]
\centering
\includegraphics[width=0.99\linewidth]{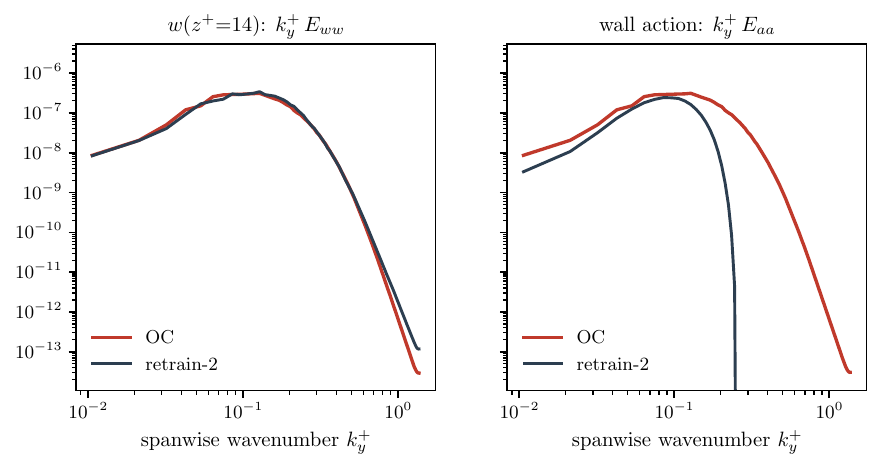}
\caption{Spanwise counterpart of Figure~\ref{fig:gap_kx}: premultiplied spanwise
spectra of the detection-plane velocity (left) and the deployed action (right).
The action is truncated beyond $k_y^{+}\sim0.26$ by the low-pass; the underlying field
sits slightly above the OC through the energetic band.}
\label{fig:gap_ky}
\end{figure}

Two spectral diagnostics show where in wavenumber space the gap
between OC and the wall-only controller lies. When considering  the streamwise
direction, Figure~\ref{fig:gap_kx} shows that the detection-plane
velocity in the twice-retrained flow exceeds the OC reference slightly across the full range of streamwise
wavenumbers. 
The wall actuation (right panel of Figure~\ref{fig:gap_kx}) is correspondingly
reduced at the energetic scales and falls off at higher wavenumbers, where the spectral constraint limits the range of
the deployed action.
In the spanwise direction, Figure~\ref{fig:gap_ky} shows that 
the wall actuation is truncated at high $k_y$ by the low-passfilter, while the underlying detection-plane field again exceeds
the OC slightly through the energetic band.
This is consistent with the two-dimensional spectra of Figures~\ref{fig:z142dww} and~\ref{fig:z142dco},
where the energetic peak lies at the OC scales but carries somewhat more energy under the wall-only control. The residual gap is
thus the direct and quantifiable consequence of excluding the high-wavenumber band that wall measurements cannot resolve; it is
small, and it falls exactly where the coherence analysis of Section~\ref{sec:why} predicts.

\section{Conclusions}\label{sec:discussion}
The aim of this work was to reproduce the classic opposition control using only
wall-based sensing, with a neural-network estimator that infers the
detection-plane velocity from the two wall-shear components.

To this end, the estimator was initially trained on opposition-control data and deployed directly in the closed loop.
However, this direct approach fails because the accuracy achieved offline does not carry over to the closed loop: the controller
carries the flow away from the statistical state on which the estimator was trained, and the significant errors arise within
a high-wavenumber band that the controlled flow amplifies rather than dissipates.
Damping the actuation amplitude alone, for instance by temporal smoothing, makes matters worse: it prevents divergence while
allowing the prediction to degrade faster, so that a simulation that remains stable may drift far from the opposition-controlled
flow without any obvious sign of failure.

Each failure serves as its own diagnosis and points to a two-parts correction. The first correction applies a spanwise spectral filter to the
wall actuation, so that its energy spectrum matches that of the flow being controlled. This eliminates the high-wavenumber content
responsible for the instability and stabilises the closed loop without restricting the actuation amplitude.

The second correction retrains the estimator on data collected during the closed-loop run itself, so that it is fitted on the
same flow it controls rather than on the opposition-control data used initially. A single retraining recovers near-OC drag
reduction but the performance still drifts over the test window; a second retraining removes the drift and yields a stable mean drag reduction of
$19.7\%$, approximately three percentage points below the standard OC value of $22.7\%$, also delivering wall-shear statistics within about one percent of
the OC values. 
The remaining three-point percentage shortfall in drag reduction falls exactly in the high-spanwise-wavenumber band where, as shown in
Section~\ref{sec:why}, the coherence $\gamma^2$ between the wall-shear signal and the detection-plane velocity drops below
the energy-containing range; in this band the wall signal carries insufficient information for any estimator to close the gap.

The wall-only controlled flow reproduces the near-wall turbulence structure of opposition control, beyond the drag reduction already
reported. One-point statistics, the quadrant and cospectral content of the Reynolds shear stress, and two-dimensional energy
spectra spanning the near-wall and detection layers all agree: the twice-retrained channel departs from the opposition-controlled
flow only by the slight near-wall excess expected of a marginally weaker controller.

The principle of training on data from the deployed flow is shared with dataset aggregation
\cite{ross2011dagger} and solver-in-the-loop learning \cite{um2020solverloop}.
Specific to this work are the diagnosis tying the closed-loop instability to a
spectral band that wall measurements cannot resolve, the spectral constraint that
makes the in-loop data collection possible, and the characterisation of the
controlled flow against the opposition-control reference.

Some questions remain open. The spectral constraint is imposed only in the
spanwise direction; a streamwise counterpart is straightforward but was not
needed for stability, and whether it would recover any of the residual gap is
unclear, since that gap lies in a band the wall signal cannot inform. We have
also not tested how the retrained controller responds to large perturbations
of its initial state, nor how far a single matching-and-retraining procedure
carries across Reynolds numbers and sensor layouts, where opposition control
itself weakens \cite{stroh2015oppositioncomparison,zhou2025drlhighre}.

One practical implication of the spectral constraint concerns experimental implementation. The spanwise spectrum of the wall
actuation required to achieve a target drag reduction (Fig.~\ref{fig:spectrum}) is a quantitative target against which
any candidate wall sensor can be assessed: it specifies which spanwise wavenumbers must be resolved and to what accuracy.
Comparing what a candidate wall sensor can deliver with what the controller requires would establish whether opposition control
from wall measurements is achievable in practice.

In summary, wall-only opposition control is achievable provided
the spectral content of the actuation is matched to the controlled
flow and the estimator is trained on data from that same flow;
offline accuracy alone is not a reliable indicator of closed-loop
performance.
%
%

\backmatter

\section*{Declarations}
\begin{itemize}
\item Funding: G.C.\ acknowledges support from the EPSRC Doctoral Training
Partnership, grant EP/W524608/1. M.P.C.\ was supported by the
Horizon Europe Marie Sk\l{}odowska-Curie Doctoral Network SCALE,
grant agreement No.\ 101120014.
\item Competing interests: the authors declare no competing interests.
\item Author contributions: G.M.C.\ developed the code, ran the simulations,
analysed the results, and conceptualised the stabilisation strategies.
M.P.C.\ contributed to conceptualising the stabilisation strategies and to
preparing the manuscript. A.P.\ contributed to preparing the manuscript and to
the conclusions. All authors read and approved the final manuscript.
\item Ethics approval: not applicable.
\item Consent to participate: not applicable.
\item Consent for publication: not applicable.
\item Data and code availability: code to reproduce the results is available at
\url{https://github.com/gmcavallazzi/CaNS_opposition_wall_only}.
\end{itemize}

\begin{appendices}

\section{Network and training details}\label{app:training}

The estimator is the encoder--ConvGRU--decoder of
Table~\ref{tab:arch}, with two-channel input $(\taux,\tauy)$,
a $64$-channel hidden state and a single scalar output $\what$
on the $256\times256$ plane. Every convolution uses a $3\times3$
kernel with circular padding, except the final $1\times1$
readout.

Training minimises the mean-squared error against the true
detection-plane velocity, which the simulation provides at every
cycle. Inputs and target are normalised to unit r.m.s.\ per
channel using the values in Table~\ref{tab:stages}, and the
target has its wall-parallel mean removed. The network is
unrolled over sequences of twelve cycles by backpropagation
through time, with the first three cycles kept as recurrent
warm-up and excluded from the loss. Optimisation uses Adam with
learning rate $5\times10^{-4}$ and weight decay $10^{-5}$;
the gradient norm is clipped at $1.0$, the batch size is four,
and the train--validation split is $90/10$. A plateau scheduler
halves the learning rate after four epochs without improvement;
each stage runs for thirty epochs.

The three stages differ only in their data and normalisation.
The first network (gen-0) is trained on approximately five
thousand cycles of converged OC. Each subsequent network is
trained on the thousand-cycle closed-loop run of the preceding
controller, under the spectral treatment of
Section~\ref{sec:fix1}, with the normalisation reset to that
run's r.m.s. The architecture, optimiser and schedule are
identical across the three stages.
\end{appendices}


\end{document}